\documentclass[12pt]{article}
\usepackage{cite}
\usepackage{float}
\usepackage{amsfonts}
\usepackage{amsmath}
\usepackage{amsbsy}
\usepackage{graphicx}
\usepackage{amssymb}
\usepackage{amsthm}
\usepackage{bm}
\usepackage{epsfig}
\usepackage{latexsym}
\usepackage{pdflscape}
\usepackage{color}
\numberwithin{equation}{section}
\input amssym.def
\input amssym.tex
\allowdisplaybreaks

\DeclareMathOperator{\tr}{tr}
\DeclareMathOperator{\Ai}{Ai}
\DeclareMathOperator{\vol}{vol}
\DeclareMathOperator{\sgn}{sgn}

\newcounter{aff}

\begin{document}

\begin{titlepage}
\begin{flushright}
{\footnotesize YITP-14-56}
\end{flushright}
\begin{center}
{\large\bf Partition Functions
of Superconformal Chern-Simons Theories from Fermi Gas Approach}

\bigskip\bigskip
{\large 
Sanefumi Moriyama\footnote[1]{\tt moriyama@math.nagoya-u.ac.jp}
\quad and \quad
Tomoki Nosaka\footnote[2]{\tt nosaka@yukawa.kyoto-u.ac.jp}
}\\
\bigskip
${}^{*}$\,
{\small\it Kobayashi Maskawa Institute
\& Graduate School of Mathematics, Nagoya University\\
Nagoya 464-8602, Japan}
\medskip\\
${}^{*\dagger}$\,
{\small\it Yukawa Institute for Theoretical Physics,
Kyoto University\\
Kyoto 606-8502, Japan}
\end{center}

\begin{abstract}
We study the partition function of three-dimensional ${\mathcal N}=4$
superconformal Chern-Simons theories of the circular quiver type,
which are natural generalizations of the ABJM theory, the worldvolume
theory of M2-branes.
In the ABJM case, it was known that the perturbative part of the
partition function sums up to the Airy function as
$Z(N)=e^{A}C^{-1/3}\Ai[C^{-1/3}(N-B)]$ with coefficients $C$, $B$ and $A$
and that for the non-perturbative part the divergences coming from the
coefficients of worldsheet instantons and membrane instantons cancel
among themselves.
We find that many of the interesting properties in the ABJM theory are
extended to the general superconformal Chern-Simons theories.
Especially, we find an explicit expression of $B$ for general
${\mathcal N}=4$ theories, a conjectural form of $A$ for a special
class of theories, and cancellation in the non-perturbative coefficients for the
simplest theory next to the ABJM theory.
\end{abstract}

\end{titlepage}
\tableofcontents

\section{Introduction and summary}\label{intro}

There is no doubt that M-theory is one of the most important achievements
in theoretical physics, though, at the same time, it is one of the
most mysterious theories.
It is a famous result from the AdS/CFT correspondence \cite{KT} that
the number of degrees of freedom of the stack of $N$ M2-branes is $N^{3/2}$ and that of the stack of M5-branes is $N^3$.
With these novel large $N$ behaviors which are in contrast to the intuitive behavior $N^2$ of D-branes, it is obvious
that these M-theoretical branes deserve intensive studies.

The M2-brane worldvolume theory on the flat spacetime was explored by
supersymmetrizing the topological Chern-Simons theory \cite{S} and
finally it was proposed \cite{ABJM} that the worldvolume theory of $N$
M2-branes on the geometry ${\mathbb C}^4/{\mathbb Z}_k$ is
described by ${\mathcal N}=6$ supersymmetric Chern-Simons theory with
gauge group $U(N)_k\times U(N)_{-k}$ and bifundamental matters between
them, which is dubbed ABJM theory.
Here the subscript $k$ and $-k$ are the Chern-Simons levels associated
to each $U(N)$ factor.

Following recent progress of localization techniques
\cite{P,KWY,J,HHL}, it was found that for the partition function and
vacuum expectation values of supersymmetric quantities in the ABJM
theory, the infinite-dimensional path integral in defining these
quantities is reduced to a finite-dimensional matrix integration.
Furthermore, due to the large supersymmetries, many interesting properties of this ABJM matrix model are discovered
\cite{DT,MPtop,DMP1,HKPT,DMP2,FHM,O,MP,KEK,KMSS,
HMO1,PY,HMO2,CM,HMO3,GKM,HMMO,HHMO,KM}:
the perturbative part of the partition function sums up to the Airy
function \cite{FHM}; the divergences in the coefficients of membrane
instantons and those of worldsheet instantons cancel among themselves
\cite{HMO2}; the non-perturbative part of the partition function is
expressed in terms of the refined topological string \cite{HMMO}.

Here we briefly review these results on the partition function of the
ABJM theory.
First let us consider the perturbative part.\footnote{
The partition function can be studied in the perturbation of $1/N$
with the 't Hooft coupling $\lambda=N/k$ fixed from the stringy regime or with
the M-theory background $k$ fixed from the M-theory regime.
The perturbation can be understood in either sense.}
It was predicted from the gravity dual that the 't Hooft
coupling $\lambda=N/k$ should be shifted as
$\lambda_{\rm eff}=\lambda-1/24+1/(24k^2)$ \cite{AHHS} and (except an
inconsistency in the coefficient of the $k^{-2}$ correction) this shift
was captured from the study of the matrix model \cite{DMP1,DMP2}.
With the shift of the 't Hooft coupling in mind, the all genus perturbative
corrections of the partition function sum up to the Airy
function \cite{FHM}
\begin{align}
Z_\text{pert}(N)=e^{A}C^{-1/3}\Ai\bigl[C^{-1/3}(N-B)\bigr],
\label{Ai}
\end{align}
using the relation with the topological string theory on local
${\mathbb P}^1\times{\mathbb P}^1$ \cite{MPtop,DMP1}.
This result was later beautifully rederived \cite{MP} by rewriting the
ABJM partition function into the partition function of a Fermi gas
system, without mentioning the relation with the topological string.
Here the $N$-independent constants $C$ and $B$ are given by simple
functions of $k$
\begin{align}
C_\text{ABJM}(k)=\frac{2}{\pi^2k},\quad
B_\text{ABJM}(k)=\frac{1}{3k}+\frac{k}{24},
\end{align}
while $A$ is a very complicated function
\begin{align}
A_\text{ABJM}(k)=-\frac{1}{6}\log\frac{k}{4\pi}+2\zeta'(-1)
-\frac{\zeta(3)}{8\pi^2}k^2+\frac{1}{3}\int\frac{dx}
{e^{kx}-1}\biggl(\frac{3}{x\sinh^2x}-\frac{3}{x^3}+\frac{1}{x}\biggr),
\label{aABJM}
\end{align}
which was obtained by taking the Borel sum of the constant map
contribution \cite{KEK}.

The non-perturbative effects have a more drastic structure.
It turns out that there are two types of non-perturbative effects.
One is called worldsheet instanton \cite{WSinst,DMP1} which
corresponds to the string worldsheet wrapping the holomorphic cycle
$\mathbb{CP}^1$ in $\mathbb{CP}^3$, while the other is called membrane instanton
\cite{DMP2} which corresponds to the D2-brane wrapping the Lagrangian
submanifold $\mathbb{RP}^3$ of $\mathbb{CP}^3$, where $\mathbb{CP}^3$ comes from the string theory limit $k\to\infty$ of $\mathbb{C}^4/\mathbb{Z}_k$.
It was found \cite{HMO2} that the coefficients of both instanton
effects are actually divergent at certain levels $k$, though the
divergences are cancelled perfectly, if we include all of the
non-perturbative effects including worldsheet instantons, membrane
instantons and their bound states.
This cancellation mechanism helps us to determine the whole
non-perturbative effects \cite{HMMO}.

It is interesting to ask whether the beautiful structures in the ABJM
theory persist in other theories.
Before arriving at the ABJM theory, a large class of supersymmetric
Chern-Simons theories were found.
For ${\mathcal N}=3$, the supersymmetric Chern-Simons theories are
constructed \cite{ZK,KL,KLL} for general gauge groups and general
matter contents.
Especially, the theory has the conformal symmetry when the gauge group
is $\prod_{a=1}^MU(N)_{k_a}$ with $\sum_{a=1}^Mk_a=0$ and the matter
fields are in the bifundamental representation of $U(N)_{k_a}$ and
$U(N)_{k_{a+1}}$ \cite{GY,JT}.
These theories can be expressed by a circular quiver taking the same
form as the Dynkin diagram of $\widehat A_{M-1}$, where each vertex
denotes the $U(N)$ factor of the gauge group and each edge denotes the
bifundamental matter.
The Chern-Simons theory with less supersymmetries is believed to
describe the worldvolume theory of multiple M2-branes on a geometry
with less supersymmetries.

\begin{figure}
\centering
\includegraphics[width=10cm]{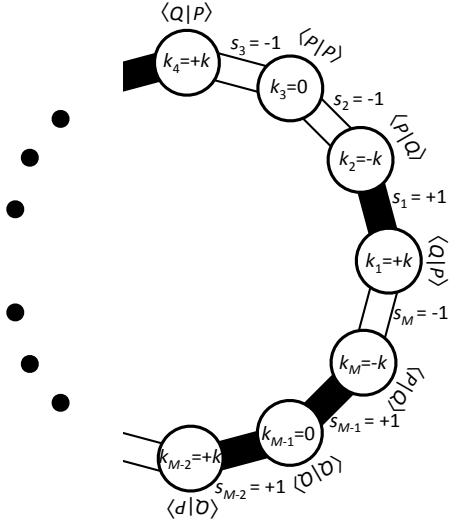}
\caption{Circular quiver with
$\{s_a\}_{a=1}^M=\{(+1),(-1)^2,\cdots,(+1)^2,(-1)\}$, which is
associated to the ${\mathcal N}=4$ superconformal Chern-Simons
theories.
Here we paint the edges assigned with $s_a=+1$ black, and those
assigned with $s_a=-1$ white.
}
\label{N4quiver}
\end{figure}

It was found that among others when the number of the $U(N)$ factors
is even and the levels are $k$ and $-k$ appearing alternatively, the
number of the supersymmetries is enhanced to ${\mathcal N}=4$
\cite{GW,HLLLP1} and the background geometry is interpreted to be an
orbifold \cite{BKKS,IK,TY}.
As pointed out in \cite{IK4}, the ${\mathcal N}=4$ enhancement is not
restricted to the case of alternating levels.
In fact, as long as the levels are expressed as
\begin{align}
k_a=\frac{k}{2}(s_a-s_{a-1}),\quad s_a=\pm 1,
\label{pm1}
\end{align}
the supersymmetry of all these theories extends to ${\mathcal N}=4$.
Since these theories are characterized by $s_a$ which
are associated to the edges of the quivers and take only two signs, it is more
suitable to assign two colors to the edges, rather than to
paint the vertices.
See figure \ref{N4quiver}.

With its simplicity in derivation, the authors of \cite{MP} were able
to further argue that, for the large class of general ${\mathcal N}=3$
superconformal circular quiver Chern-Simons theories (associated with
a hermitian Hamiltonian, as explained later), the perturbative
partition function is always given by the same form \eqref{Ai} with
some coefficients $C$, $B$ and $A$.
Also, the study of the large $N$ behavior (the coefficient $C$) in
many theories can be found in earlier works
\cite{SMP,MS,CKK,JKPS,GHP,S1,GAH,GHN,
interacting,S2,CHJ,S3,MePu,GM,AZ}.
Especially it is worthwhile to mention that, according to \cite{MP}
the expression of the coefficient $C$ was given a geometrical
interpretation as the classical volume inside the Fermi surface.
Moreover, recently in \cite{HM} the special ${\mathcal N}=4$ case with
the gauge group $[U(N)_k\times U(N)_{-k}]^r$, whose quiver is the
$r$-ple repetition of that of the ABJM theory, was studied carefully
including the instanton effect using the relation to the original ABJM
theory. 
Alongside, the authors found that if the circular quiver is the
$r$-ple repetition of a ``fundamental" circular quiver, the grand
potential of the repetitive theory is given explicitly by that of the
``fundamental" theory.
Especially, it was found that the perturbative coefficients of the
$r$-ple repetitive theory $[C]_r$, $[B]_r$, $[A]_r$ are related to
$[C]_1$, $[B]_1$, $[A]_1$ by
\begin{align}
[C]_r=\frac{1}{r^2}[C]_1,\quad
[B]_r=[B]_1-\frac{\pi^2}{3}\biggl(1-\frac{1}{r^2}\biggr)[C]_1,\quad
[A]_r=r[A]_1.
\label{CBAr}
\end{align}
However, the coefficients $B$, $A$ and the non-perturbative
corrections for general ${\cal N}=3$ theories have not been known so
far.

In this paper, we extend the previous studies on the ABJM theory to
the more general ${\mathcal N}=4$ cases with the levels \eqref{pm1}.
Especially we hope that after figuring out the cancellation mechanism
among all of the instanton effects, the instanton moduli space of the
membrane theories will become clearer.
We first concentrate on the perturbative part.
Using the Fermi gas formalism, we give an explicit formula for $B$,
which is deeply related to the redefinition of the 't Hooft coupling.
We have found that, when the edges are assigned with
\begin{align}
\{s_a\}_{a=1}^M=\{(+1)^{q_1},(-1)^{p_1},(+1)^{q_2},(-1)^{p_2},
\cdots,(+1)^{q_m},(-1)^{p_m}\},
\label{sqp}
\end{align}
the coefficient $B$ is given by
\begin{align}
B&=\frac{B^{(0)}}{k}+kB^{(2)},
\label{B}
\end{align}
with
\begin{align}
B^{(0)}&=-\frac{1}{6}\biggl[
\frac{\Sigma(p)}{\Sigma(q)}+\frac{\Sigma(q)}{\Sigma(p)}
-\frac{4}{\Sigma(q)\Sigma(p)}\biggr],\nonumber\\
B^{(2)}&=\frac{1}{24}\biggl[\Sigma(q)\Sigma(p)
-12\biggl(\frac{\Sigma(q,p,q)}{\Sigma(q)}
+\frac{\Sigma(p,q,p)}{\Sigma(p)}
-\frac{\Sigma(q,p)\Sigma(p,q)}{\Sigma(q)\Sigma(p)}\biggr)\biggr].
\label{B0B2}
\end{align}
Here we adopt the notation of $\Sigma(L)$, with $L$ denoting an
alternating sequence of $q$ and $p$, whose definition is given by
\begin{align}
&\Sigma(q)=\sum_{a=1}^mq_a,\quad
\Sigma(p)=\sum_{a=1}^mp_a,\nonumber\\
&\Sigma(q,p)=\sum_{1\le a\le b\le m}q_ap_b,\quad
\Sigma(p,q)=\sum_{1\le a<b\le m}p_aq_b,\nonumber\\
&\Sigma(q,p,q)=\sum_{1\le a\le b<c\le m}q_ap_bq_c,\quad
\Sigma(p,q,p)=\sum_{1\le a<b\le c\le m}p_aq_bp_c.\label{Sigmadef}
\end{align}
Note that the condition in each summation can be restated as the
requirement that we choose $q_a$'s and $p_a$'s out of
$q_1,p_1,q_2,p_2,\cdots,q_m,p_m$ by respecting its ordering.
We stress that the result \eqref{B} with \eqref{B0B2} is encoded
suitably in this notation.
After we introduce this notation, the proof of the result is quite
straightforward.

It is still difficult to obtain the general expression of the
coefficient $A$ with the current technology.
For the special case when the edges of $s_a=+1$ and those of $s_a=-1$
are clearly separated
\begin{align}
\{s_a\}_{a=1}^M=\{(+1)^q,(-1)^p\},
\label{separate}
\end{align}
we conjecture that the coefficient $A$ is given in terms of the
coefficient of the ABJM case $A_{\rm ABJM}(k)$ \eqref{aABJM} by
\begin{align}
A(k)=\frac{1}{2}\bigl(p^2A_{\rm ABJM}(qk)+q^2A_{\rm ABJM}(pk)\bigr).
\label{A}
\end{align}
Later we shall provide evidences for this conjecture using the WKB
expansion \eqref{Awkb} and numerical data (table \ref{fit}).

After determining the perturbative part, we continue to the
non-perturbative part.\footnote{The interpretation of these
non-perturbative instanton effects in the gravity dual still awaits to
be studied carefully. In this paper we call these non-perturbative
instanton effects membrane instanton when the exponent is proportional
to $\mu$ while we call them worldsheet instanton when the exponent is
proportional to $\mu/k$.}
To fully understand the non-perturbative instanton effects, we still
need lots of future studies.
We shall concentrate on the separative case \eqref{separate} with
$q=2$, $p=1$, that is, $\{s_a\}_{a=1}^3=\{(+1)^2,(-1)\}$, which is the
simplest case other than the ABJM theory.
Using the WKB expansion of the Fermi gas formalism, we can study the
membrane instanton order by order in $\hbar=2\pi k$.
We have found that the first membrane instanton is consistent with
\begin{align}
J^\text{MB}_\text{np}(\mu)=-\frac{2}{\tan\frac{\pi k}{2}}e^{-\mu}
+{\mathcal O}(e^{-2\mu}),
\label{c1}
\end{align}
up to the ${\mathcal O}(k^5)$ term in the $\hbar=2\pi k$ expansion.
On the other hand, using the numerical coefficients of the grand potential for
$k=3,4,5,6$, we conjecture that the first worldsheet instanton is
given by
\begin{align}
J^\text{WS}_\text{np}(\mu)
=\frac{4\cos\frac{\pi}{k}}{\sin^2\frac{2\pi}{k}}e^{-\frac{2}{k}\mu}
+{\mathcal O}(e^{-\frac{4}{k}\mu}).
\label{d1}
\end{align}
We can see that the coefficients of both the first membrane instanton
\eqref{d1} and the first worldsheet instanton \eqref{c1} are divergent
at $k=2$ and the remaining finite part after cancelling the
divergences matches perfectly with the numerical coefficients at $k=2$.

The remaining part of this paper is organized as follows.
In section \ref{FermiGas}, we shall demonstrate the Fermi gas
formalism for general ${\mathcal N}=4$ superconformal Chern-Simons
theories.
Then in section \ref{secFS} we shall proceed to derive the expression
of $B$ for general ${\mathcal N}=4$ circular quivers.
We shall shortly see the consistency with the transformation under the
repetition in section \ref{repetition} and see the possible
generalization to the ${\mathcal N}=3$ cases in section \ref{N3}.
After that, we shall turn to the WKB expansion of the grand potential
in section \ref{AandJnp}, where not only the consistency with the
expression of $B$ but also further information on the coefficient $A$
and the instantons are found.
In section \ref{CancelMech} we shall study the non-perturbative
instanton effects for the special case of
$\{s_a\}_{a=1}^3=\{(+1)^2,(-1)\}$.
Finally in section \ref{Discussion} we conclude with some future
directions.

{\bf Note added.}
As our work had been completed and we were in the final stage of checking the draft, Hatsuda and Okuyama submitted their work \cite{HaOk}, where they also used the Fermi gas formalism to study the $N_f$ matrix model \cite{GM}.
Although the original theories are different, in terms of the Fermi gas formalism, the density matrix (2.4) in \cite{HaOk} is reproduced if we restrict our setup to the separative case $\{s_a\}_{a=1}^M=\{(+1)^{N_f},(-1)\}$ and put $k=1$.
Their results also have some overlaps with ours.
For example, our conjectural form of the coefficient $A$ \eqref{A} reduces to their conjecture (3.12) in \cite{HaOk} under this restriction.

\section{${\mathcal N}=4$ Chern-Simons matrix model as a Fermi gas}
\label{FermiGas}

In this section we shall show that the partition functions of
${\mathcal N}=4$ superconformal circular quiver Chern-Simons theories,
with gauge group $\prod_{a=1}^MU(N)_{k_a}$ and Chern-Simons levels
chosen to be \eqref{pm1}, can be regarded as the partition functions of
$N$-particle ideal Fermi gas systems governed by non-trivial
Hamiltonians.
Although this structure was already proved in \cite{MP} for more
general ${\mathcal N}=3$ superconformal circular quiver Chern-Simons
theories without the restriction of levels \eqref{pm1}, we shall
repeat the derivation since the special simplification occurs for
${\mathcal N}=4$ theories with the levels \eqref{pm1}.
In particular we find that, corresponding to the colors of edges
$\{s_a\}_{a=1}^M$ \eqref{sqp}, the Hamiltonian of the associated Fermi
gas system is given by
\begin{align}
e^{-{\widehat H}}
&=\biggl[2\cosh\frac{\widehat Q}{2}\biggr]^{-q_1}
\biggl[2\cosh\frac{\widehat P}{2}\biggr]^{-p_1}
\cdots
\biggl[2\cosh\frac{\widehat Q}{2}\biggr]^{-q_m}
\biggl[2\cosh\frac{\widehat P}{2}\biggr]^{-p_m}.
\label{Hamiltonian}
\end{align} 

Let us begin with the partition function of an ${\mathcal N}=4$ circular
quiver Chern-Simons theory with gauge group $[U(N)]^M$ and levels
\eqref{pm1},
\begin{align}
Z(N)=\frac{1}{(N!)^M}
\int\Biggl(\prod_{a=1}^M\prod_{i=1}^ND\lambda_{a,(i)}\Biggr)
\left(\prod_{a=1}^M\frac{\prod_{i<j}^N
\left(2\sinh\frac{\lambda_{a,(i)}-\lambda_{a,(j)}}{2}\right)^2}
{\prod_{i,j}^N2
\cosh\frac{\lambda_{a+1,(i)}-\lambda_{a,(j)}}{2}}\right),
\label{CSMM}
\end{align}
obtained by localization technique \cite{KWY}.
Here $M$ is the number of vertices and the integration measure is given by
\begin{align}
D\lambda_{a,(i)}=\frac{d\lambda_{a,(i)}}{2\pi}
\exp\biggl({\frac{ik_a}{4\pi}\lambda_{a,(i)}^2}\biggr),
\end{align}
with $k_a$ being the Chern-Simons level for the $a$-th $U(N)$ factor
of the gauge group $[U(N)]^M$.

Using the Cauchy identity
\begin{align}
\frac{\prod_{i<j}^N(x_i-x_j)\prod_{i<j}^N(y_i-y_j)}
{\prod_{i,j}(x_i+y_j)}
=\det\!{}_{i,j}\frac{1}{x_i+y_j},
\end{align}
and the integration formula \cite{MM}
\begin{align}
\frac{1}{N!}\int\biggl(\prod Dx_k\biggr)
\det\!{}_{i,k}(\phi_i(x_k))\det\!{}_{j,k}(\psi_j(x_k))
=\det\!{}_{i,j}\biggl(\int Dy\phi_i(y)\psi_j(y)\biggr),
\end{align}
we find that the partition function is
\begin{align}
Z(N)=\frac{1}{N!}\int\Biggl(\prod_{i=1}^N D\lambda_{1,(i)}\Biggr)
\sum_{\sigma\in S_N}(-1)^\sigma
\rho(\lambda_{1,(\sigma(i))},\lambda_{1,(i)}),
\label{ZN}
\end{align}
where the density matrix $\rho(x,y)$ is given by
\begin{align}
\rho(x,y)=\int\Biggl(\prod_{a=2}^{M}D\lambda_a\Biggr)
\frac{1}{2\cosh\frac{x-\lambda_M}{2}}
\Biggl(\prod_{a=2}^{M-1}
\frac{1}{2\cosh\frac{\lambda_{a+1}-\lambda_{a}}{2}}\Biggr)
\frac{1}{2\cosh\frac{\lambda_2-y}{2}}.
\label{rho}
\end{align}
If we introduce the grand potential $J(\mu)$ as
\begin{align}
e^{J(\mu)}=1+\sum_{N=1}^\infty e^{\mu N}Z(N),\label{ZtoJ}
\end{align}
the sum over the permutation in \eqref{ZN} simplifies into
\begin{align}
J(\mu)=\tr\log(1+e^\mu\rho).
\end{align}
Here both the multiplication among $\rho$ and the trace are performed
with $D\lambda_1$, just as the multiplication within $\rho$
\eqref{rho} which is performed with $D\lambda_a$ $(a=2,\cdots,M)$.
Introducing the Fourier transformation $(\lambda_{M+1}=\lambda_1)$
\begin{align}
\frac{1}{2\cosh\frac{\lambda_{a+1}-\lambda_a}{2}}
=\int\frac{d\Lambda_a}{2\pi}
\frac{e^{\frac{i}{2\pi}(\lambda_{a+1}-\lambda_a)\Lambda_a}}
{2\cosh\frac{\Lambda_a}{2}},
\end{align}
for all $a$, we find that the integration associated with $\lambda_a$
in $\tr\rho^n$ is given by
\begin{align}
\int\cdots\frac{d\Lambda_a}{2\pi}\frac{1}{2\cosh\frac{\Lambda_a}{2}}
\frac{d\Lambda_{a-1}}{2\pi}\frac{1}{2\cosh\frac{\Lambda_{a-1}}{2}}\cdots
\int\frac{d\lambda_a}{2\pi}
\exp\biggl[\frac{ik_a\lambda_a^2}{4\pi}
-\frac{i(\Lambda_{a}-\Lambda_{a-1})\lambda_a}{2\pi}\biggr].
\end{align}
If we introduce the coordinate variables $\Lambda_a=Q_a$ for $s_a=+1$
and the momentum variables $\Lambda_a=P_a$ for $s_a=-1$, we find that,
up to an irrelevant numerical factor which will be cancelled out
finally, this integration essentially gives that
\begin{align}
\int\frac{d\lambda_a}{2\pi}
\exp\biggl[\frac{ik_a\lambda_a^2}{4\pi}
-\frac{i(\Lambda_{a}-\Lambda_{a-1})\lambda_a}{2\pi}\biggr]
\simeq \langle\Lambda_a|\Lambda_{a-1}\rangle,
\end{align}
because the inner products of the coordinate and momentum eigenstates
are given by
\begin{align}
&\langle Q_a|Q_{a-1}\rangle=2\pi\delta(Q_a-Q_{a-1}),\quad
\langle P_a|P_{a-1}\rangle=2\pi\delta(P_a-P_{a-1}),\nonumber\\
&\langle Q_a|P_{a-1}\rangle=\frac{1}{\sqrt{k}}e^{iQ_aP_{a-1}/(2\pi k)},
\quad
\langle P_a|Q_{a-1}\rangle=\frac{1}{\sqrt{k}}e^{-iP_aQ_{a-1}/(2\pi k)}.
\end{align}
Finally the integration in $\tr\rho^n$ is given by
\begin{align}
\int\cdots\frac{d\Lambda_a}{2\pi}
\frac{d\Lambda_{a-1}}{2\pi}
\cdots
\frac{1}{2\cosh\frac{\Lambda_a}{2}}\langle\Lambda_a|\Lambda_{a-1}\rangle
\frac{1}{2\cosh\frac{\Lambda_{a-1}}{2}}\cdots.
\end{align}

This means that, if we define the position and momentum operator
${\widehat Q},{\widehat P}$ obeying the canonical commutation relation
\begin{align}
[{\widehat Q},{\widehat P}]=i\hbar,
\end{align}
with $\hbar=2\pi k$, the Hamiltonian
${\widehat H}({\widehat Q},{\widehat P})$ is given as
\eqref{Hamiltonian} for the ordering \eqref{sqp} 
(see figure \ref{N4quiver}).
Therefore, the grand potential $J(\mu)$ can be interpreted as the
grand potential of the ideal Fermi gas system with $N$ particles whose
one-particle Hamiltonian ${\widehat H}$ is given by
\eqref{Hamiltonian}, where $\mu$ is the chemical potential dual to the
number of particles $N$.

\section{Fermi surface analysis}\label{secFS}
In the previous section we have constructed the Fermi gas formalism
for ${\cal N}=4$ superconformal Chern-Simons theories by rewriting the
partition function into that of non-interacting $N$-particle Fermi gas
systems with non-trivial Hamiltonians \eqref{Hamiltonian}.

Note that the Hamiltonian \eqref{Hamiltonian} is non-hermitian.
In some particular cases, including the ABJM theory, however, we can
choose it to be hermitian by redefining the Hamiltonian by
\begin{align}
e^{-{\widehat H}}\rightarrow
\biggl(2\cosh\frac{\widehat Q}{2}\biggr)^{x}
e^{-{\widehat H}}
\biggl(2\cosh\frac{\widehat Q}{2}\biggr)^{-x},\label{hermitianization}
\end{align}
with a real number $x$, which does not affect the trace.
Below, we shall restrict ourselves to these cases.

It was argued in \cite{MP} that, for a large class of general
${\mathcal N}=3$ superconformal circular quiver Chern-Simons theories
associated to a hermitian Hamiltonian ${\widehat{H}}$ in the above
sense, the number $n(E)$ of states whose eigenvalue of ${\widehat{H}}$
is smaller than $E$ is universally given as
\begin{align}
n(E)=CE^2+n(0)+\text{non-pert},
\label{nEfinal}
\end{align}
with $C$ and $n(0)$ being constants depending on $k$ and
``$\text{non-pert}$'' standing for non-perturbative corrections.
From this form the authors showed that the perturbative part of the
grand potential is given by a cubic potential
\begin{align}
J_\text{pert}(\mu)=\frac{C}{3}\mu^3+B\mu+A,
\end{align}
where the coefficient $B$ is given by
\begin{align}
B=n(0)+\frac{\pi^2C}{3}.
\label{Bn0}
\end{align}
However, the explicit forms of $n(0)$ and $A$ for the general circular
quivers were not known.

In this section we shall calculate $n(0)$ and $C$ explicitly for the
class of ${\mathcal N}=4$ superconformal circular quiver Chern-Simons
theories, from the study of the Fermi surface as in \cite{MP}.
The results are, for the quiver \eqref{sqp},
\begin{align}
C&=\frac{2}{\pi^2k\Sigma(q)\Sigma(p)},\label{C}\\
n(0)&=-\frac{1}{6k}\biggl(\frac{\Sigma(p)}{\Sigma(q)}
+\frac{\Sigma(q)}{\Sigma(p)}\biggr)+kB^{(2)},
\label{n0}
\end{align}
where $B^{(2)}$ is defined in \eqref{B0B2}.
Using \eqref{Bn0} we can read off the expression of $B$ \eqref{B}
directly from this result.

\subsection{The strategy}
We follow the strategy of \cite{MP} in calculation.
The concrete definition of the number of states $n(E)$ is
\begin{align}
n(E)=\tr\theta(E-{\widehat H}).
\label{nE}
\end{align}
If we introduce the Wigner transformation
${\widehat A}\rightarrow ({\widehat A})_{\rm W}$ with
\begin{align}
({\widehat A})_{\rm W}=\int\frac{dQ^\prime}{2\pi}
\biggl\langle Q-\frac{Q^\prime}{2}
\,\biggl|\,{\widehat A}\,\biggr|\,
Q+\frac{Q^\prime}{2}\biggr\rangle
e^{\frac{iQ^\prime P}{\hbar}},
\label{Wigner}
\end{align}
similarly to the case of ABJM theory \cite{MP}, $n(E)$ is approximated
by
\begin{align}
n(E)\simeq\int\frac{dQdP}{2\pi\hbar}\theta(E-H_{\rm W}),
\label{errortheta}
\end{align}
up to non-perturbative corrections in $E$ for large $E$.
Here we have introduced the abbreviation
$H_{\rm W}=({\widehat H})_{\rm W}$.
This means that, up to the non-perturbative corrections, $n(E)$ is
given by the volume inside the Fermi surface of the semiclassical
Wigner Hamiltonian,
\begin{align}
n(E)\simeq\frac{1}{2\pi\hbar}
\vol\{(Q,P)\in\mathbb{R}^2|H_{\rm W}(Q,P)\le E\}.
\label{Fermivolume}
\end{align}
Here $H_{\rm W}$ is calculated from \eqref{Hamiltonian} by using the
following property of the Wigner transformation
\begin{align}
({\widehat A}{\widehat B})_{\rm W}
=({\widehat A})_{\rm W}\star({\widehat B})_{\rm W},
\end{align}
with the star product given by
\begin{align}
\star=\exp\biggl[\frac{i\hbar}{2}\Bigl(
\overleftarrow{\partial}_Q\overrightarrow{\partial}_P
-\overleftarrow{\partial}_P\overrightarrow{\partial}_Q\Bigr)\biggr],
\end{align}
which follows from the definition of the Wigner transformation
\eqref{Wigner}.

Before going on, we shall argue some general properties of the Fermi
surface \eqref{Fermivolume}.
The Wigner Hamiltonian $H_{\rm W}$ obtained from the quantum
Hamiltonian \eqref{Hamiltonian} is a sum of the classical part
\begin{align}
H_{\rm W}^{(0)}=\Sigma(q)U+\Sigma(p)T,
\label{HW0}
\end{align}
with
\begin{align}
U(Q)=\log 2\cosh\frac{Q}{2},\quad T(P)=\log 2\cosh\frac{P}{2},
\label{defUT}
\end{align}
and $\hbar$-corrections which consist of derivatives of $U$ and $T$.
Also, from the behavior of $U(Q)$ and $T(P)$ in the limit of
$|Q|\rightarrow\infty$ and $|P|\rightarrow\infty$,
\begin{align}
&U=\frac{|Q|}{2}+{\mathcal O}(e^{-|Q|}),\quad
U^\prime=\frac{\sgn(Q)}{2}+{\mathcal O}(e^{-|Q|}),\quad
U^{\prime\prime}={\mathcal O}(e^{-|Q|}),\nonumber\\
&T=\frac{|P|}{2}+{\mathcal O}(e^{-|P|}),\quad
T^\prime=\frac{\sgn(P)}{2}+{\mathcal O}(e^{-|P|}),\quad
T^{\prime\prime}={\mathcal O}(e^{-|P|}),
\label{approx}
\end{align}
it follows that the Fermi surface is approaching to
\begin{align}
\Sigma(q)|Q|+\Sigma(p)|P|=2E,
\label{diamond}
\end{align}
as $E\to\infty$.

\begin{figure}
\centering
\includegraphics[width=8cm]{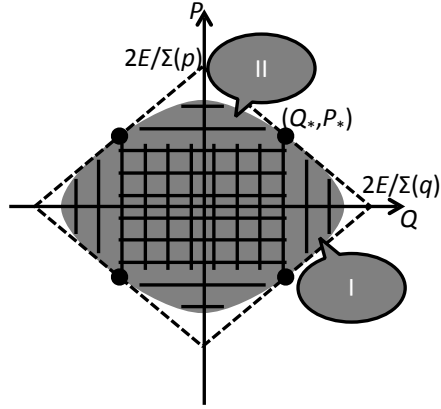}
\caption{
The Fermi surface of the ${\mathcal N}=4$ superconformal circular quiver Chern-Simons theory.
We depict region I $(|P|\le P_*)$ by the region shaded by the vertical lines, while
region II $(|Q|\le Q_*)$ denotes that shaded by the horizontal lines.}
\label{figFS}
\end{figure}

From this property, if we choose a point
$(Q_*,P_*)$ on the Fermi surface which is distant only by ${\mathcal O}(e^{-E})$ from the midpoint $(E/\Sigma(q),E/\Sigma(p))$ of the edge of \eqref{diamond}, 
the
total volume inside the Fermi surface is decomposed as
\begin{align}
\vol=\vol(\text{I})+\vol(\text{II})-2Q_*\cdot 2P_*,
\label{vtot}
\end{align}
where region I denotes the $|P|\le P_*$ part inside the Fermi surface
while region II denotes the $|Q|\le Q_*$ part.
See figure \ref{figFS}.

\subsection{Semiclassical Wigner Hamiltonian}
Now let us start concrete calculations.
The quantum Hamiltonian \eqref{Hamiltonian} is
\begin{align}
e^{-{\widehat H}}=e^{-(q_1-x){\widehat U}}e^{-p_1{\widehat T}}
e^{-q_2{\widehat U}}e^{-p_2{\widehat T}}\cdots e^{-x{\widehat U}},
\label{qH}
\end{align}
where ${\widehat U}=U({\widehat Q})$ and
${\widehat T}=T({\widehat P})$.
Here we have introduced a constant $x$ deliberately, which does not
change the trace of operators, to make ${\widehat H}$ hermitian.
Let us compute this Hamiltonian using the Baker-Campbell-Hausdorff
formula,
\begin{align}
e^Xe^Y=\exp\biggl[X+Y+\frac{1}{2}[X,Y]
+\frac{1}{12}[X,[X,Y]]+\frac{1}{12}[Y,[Y,X]]+\cdots\biggr].
\label{bch}
\end{align}

For the computation, we prepare the following formula which holds up
to higher brackets\footnote{
One can prove the formula \eqref{bchq} by induction along with its
``dual'' formula
\begin{align}
&e^{-q_1{\widehat U}}e^{-p_1{\widehat T}}
e^{-q_2{\widehat U}}e^{-p_2{\widehat T}}\cdots
e^{-p_m{\widehat T}}\nonumber\\
&=\exp\biggl[-\Sigma_m(q){\widehat U}
-\Sigma_m(p){\widehat T}
+\biggl(\Sigma_m(q,p)-\frac{1}{2}\Sigma_m(q)\Sigma_m(p)\biggr)
[{\widehat U},{\widehat T}]\nonumber\\
&\quad-\frac{1}{2}\biggl(\Sigma_m(q,p,q)
-\frac{1}{6}\Sigma_m(q)^2\Sigma_m(p)\biggr)
[{\widehat U},[{\widehat T},{\widehat U}]]
\nonumber\\
&\quad
-\frac{1}{2}\biggl(\Sigma_m(p,q,p)
-\frac{1}{6}\Sigma_m(q)\Sigma_m(p)^2\biggr)
[{\widehat T},[{\widehat U},{\widehat T}]]\biggr],
\label{bchp}
\end{align}
up to higher brackets. 
Multiplying $e^{-q_{m+1}{\widehat U}}$ to this from the right and applying
the Baker-Campbell-Hausdorff formula \eqref{bch}, one obtains the
relation \eqref{bchq}.
Also, multiplying $e^{-p_{m+1}{\widehat T}}$ from the right further, and
using an identity
\begin{align}
\Sigma_m(q,p)+\Sigma_{m(+1)}(p,q)=\Sigma_{m(+1)}(q)\Sigma_m(p),
\label{sigmasum}
\end{align}
one obtains the relation \eqref{bchp} with $m$ replaced with $m+1$.
Combining with the fact that both of the formulae hold for $m=1$, we
complete the proof of both formulae by induction.
}
\begin{align}
&e^{-q_1{\widehat U}}e^{-p_1{\widehat T}}
e^{-q_2{\widehat U}}e^{-p_2{\widehat T}}\cdots
e^{-q_{m+1}{\widehat U}}\nonumber\\
&=\exp\biggl[-\Sigma_{m+1}(q){\widehat U}-\Sigma_m(p){\widehat T}
+\biggl(\Sigma_m(q,p)-\frac{1}{2}\Sigma_{m+1}(q)\Sigma_m(p)\biggr)
[{\widehat U},{\widehat T}]\nonumber\\
&\quad-\frac{1}{2}\biggl(\Sigma_{m+1}(q,p,q)
-\frac{1}{6}\Sigma_{m+1}(q)^2\Sigma_m(p)\biggr)
[{\widehat U},[{\widehat T},{\widehat U}]]
\nonumber\\
&\quad
-\frac{1}{2}\biggl(\Sigma_m(p,q,p)
-\frac{1}{6}\Sigma_{m+1}(q)\Sigma_m(p)^2\biggr)
[{\widehat T},[{\widehat U},{\widehat T}]]\biggr],
\label{bchq}
\end{align}
and substitute $q_1-x$ into $q_1$ and $x$ into $q_{m+1}$.
Here we write explicitly the index $m$ in the definition of
$\Sigma(L)$ in \eqref{Sigmadef} to avoid confusion.
As we shall see below, higher brackets are irrelevant to the
perturbative coefficients $C$ and $B$.

We shall choose $x$ to be
\begin{align}
x=\frac{\Sigma_m(q,p)}{\Sigma_m(p)}-\frac{\Sigma_m(q)}{2},
\end{align}
so that the coefficient of a non-hermitian operator
$[{\widehat U},{\widehat T}]$ vanishes
\begin{align}
\Sigma_m(q,p)-\frac{1}{2}\Sigma_{m+1}(q)\Sigma_m(p)
\bigg|_{\begin{subarray}{c}q_1\to q_1-x\\q_{m+1}\to x
\end{subarray}}
=\Sigma_m(q,p)-x\Sigma_m(p)-\frac{1}{2}\Sigma_m(q)\Sigma_m(p)=0.
\end{align}
Then, the coefficients of
$[{\widehat U},[{\widehat T},{\widehat U}]]$ and
$[{\widehat T},[{\widehat U},{\widehat T}]]$ become
\begin{align}
&c^T=-\frac{1}{2}\biggl(\Sigma_{m+1}(q,p,q)
-\frac{1}{6}\Sigma_{m+1}(q)^2\Sigma_m(p)\biggr)\bigg|
_{\begin{subarray}{c}q_1\to q_1-x\\q_{m+1}\to x\end{subarray}}
\nonumber\\
&\quad=-\frac{1}{2}\biggl(\Sigma_m(q,p,q)
-x\Sigma_m(p,q)+x\Sigma_m(q,p)
-x^2\Sigma_m(p)-\frac{1}{6}\Sigma_{m}(q)^2\Sigma_m(p)\biggr)
\nonumber\\
&\quad=-\frac{1}{2}\biggl(\Sigma_m(q,p,q)
+\frac{1}{12}\Sigma_m(q)^2\Sigma_m(p)
-\frac{\Sigma_m(q,p)\Sigma_m(p,q)}{\Sigma_m(p)}\biggr),
\label{cT}
\end{align}
and
\begin{align}
c^U&=-\frac{1}{2}\biggl(\Sigma_m(p,q,p)
-\frac{1}{6}\Sigma_{m+1}(q)\Sigma_m(p)^2\biggr)\bigg|
_{\begin{subarray}{c}q_1\to q_1-x\\q_{m+1}\to x\end{subarray}}
\nonumber\\
&=-\frac{1}{2}\biggl(\Sigma_m(p,q,p)
-\frac{1}{6}\Sigma_{m}(q)\Sigma_m(p)^2\biggr),
\end{align}
where we have used \eqref{sigmasum} in the computation.

The Wigner Hamiltonian $H_{\rm W}$ is obtained by replacing the
operators ${\widehat U}$, ${\widehat T}$ with the functions $U$, $T$
and the operator product with the $\star$-product.
Then, we find that the $\hbar$-expansion of the Wigner Hamiltonian
\begin{align}
H_{\rm W}=\sum_{s=0}^\infty\hbar^{s}H_{\rm W}^{(s)},\label{HsubW}
\end{align}
is given by
$H_{\rm W}^{(0)}$ in \eqref{HW0} and
\begin{align}
H_{\rm W}^{(2)}&=-c^T(U')^2T''-c^U(T')^2U'',
\end{align}
up to higher order terms.
The higher order terms in $\hbar$ in \eqref{HsubW} comes from both
higher brackets and higher derivatives from the $\star$-products.
General form of such terms is
\begin{align}
\sum_{n\ge 3}\biggl[c^T_n(U^\prime)^nT^{(n)}
+c^U_n(T^\prime)^nU^{(n)}\biggr]
+\sum_{m,n\ge 2}U^{(m)}T^{(n)}(\cdots),
\label{higher}
\end{align}
with $c^T_n$ and $c^U_n$ being some constants.
Since $(Q,P)$ on the Fermi surface always satisfies either $|Q|\ge Q_*$
or $|P|\ge P_*$, the third terms are always non-perturbative according
to the asymptotic behavior of $U$ and $T$ in \eqref{approx}.
As we see below, the first two terms do not affect the volume \eqref{vtot}
up to non-perturbative corrections either.

\subsection{Volume inside the Fermi surface}
Now that the Wigner Hamiltonian with quantum corrections is obtained
to the required order, let us calculate the volume inside the Fermi
surface \eqref{Fermivolume}, following the decomposition
\eqref{vtot}.
First we consider the region I.
Since $|Q|\ge Q_*\sim E$ holds for the parts of the Fermi surface
surrounding this region, we can use the approximation \eqref{approx}
for $U$.
Then the points on the Fermi surface $H_{\rm W}=E$ are parametrized as $(Q_\pm(P),P)$ with
\begin{align}
Q_\pm(P)=\pm\frac{2}{\Sigma_m(q)}\biggl[E-\Sigma_m(p)T
+\frac{\hbar^2}{4}c^TT''
-\sum_{n\ge 3}\Bigl(\pm\frac{1}{2}\Bigr)^nc^T_nT^{(n)}\biggr]+\text{non-pert}\,,
\end{align}
with which the volume of region I is
\begin{align}
\vol(\text{I})&=\int_{-P_*}^{P_*}dP\int_{Q_-(P)}^{Q_+(P)}dQ
\nonumber\\&
=\frac{4}{\Sigma_m(q)}\biggl[2EP_*
-\Sigma_m(p)\biggl(\frac{P_*^2}{2}+\frac{\pi^2}{6}\biggr)
+\frac{\hbar^2}{4}c^T\biggr]+\text{non-pert}\,.
\end{align}
The contribution from $T^{(n)}$ with $n\ge 3$ is the surface term
$T^{(n-1)}$, which just gives non-perturbative effects due to
\eqref{approx} when evaluated at $P=\pm P_*$.
Similarly, the volume of region II is evaluated, using the
approximation \eqref{approx} for $T(P)$, as
\begin{align}
\vol(\text{II})=\frac{4}{\Sigma_m(p)}\biggl[2EQ_*
-\Sigma_m(q)\biggl(\frac{Q_*^2}{2}+\frac{\pi^2}{6}\biggr)
+\frac{\hbar^2}{4}c^U\biggr]+\text{non-pert}\,.
\end{align}

Summing up all the contributions to \eqref{vtot}, one obtains the
total volume.
After substituting the volume into \eqref{Fermivolume}, the number of states $n(E)$ is written as \eqref{nEfinal},
with $C$ and $n(0)$ given by \eqref{C} and \eqref{n0}.

\section{Repetition invariance}\label{repetition}

As explained in \eqref{CBAr}, it was found in \cite{HM} that, if the
circular quiver is the $r$-ple repetition of another fundamental circular
quiver, the coefficients $C$, $B$ and $A$ of the repetitive theory are
related to those of the fundamental theory.
This implies that the quantity $n(0)$ \eqref{Bn0} is invariant under
repetition,
\begin{align}
[n(0)]_r=[n(0)]_1.
\end{align}
In this section we show this property explicitly for the result
\eqref{n0} we have obtained in the previous section for general
${\mathcal N}=4$ circular quivers.

Suppose that the circular quiver \eqref{sqp} is the $r$-ple repetition
of a fundamental circular quiver $(M=r\tilde{M}$, $m=r\tilde{m})$
\begin{align}
\{s_a\}_{a=1}^{\tilde{M}}=\{(+1)^{\tilde{q}_1},(-1)^{\tilde{p}_1},\cdots,
(+1)^{\tilde{q}_{\tilde{m}}},(-1)^{\tilde{p}_{\tilde{m}}}\}.
\end{align}

To study how $n(0)$ changes under the repetition, let us first
consider its building block $\Sigma_m(L)$ defined in
\eqref{Sigmadef}.
For this purpose, we shall decompose the label $a$ of $q_a$ and $p_a$
into two integers $(\alpha,\tilde{a})$ by
\begin{align}
a=(\alpha-1)\tilde{m}+\tilde{a},
\end{align}
with $1\le\alpha\le r$ and $1\le\tilde{a}\le\tilde{m}$, which implies
\begin{align}
q_a=\tilde{q}_{\tilde{a}},\quad p_a=\tilde{p}_{\tilde{a}},
\end{align}
Then we find that the relation $a<b$ (or $a\le b$)
appearing in the summation in
\eqref{Sigmadef} is represented as
\begin{align}
\text{``}\alpha<\beta\text{''},
\qquad\quad\text{or}\qquad\quad
\text{``}\alpha=\beta\quad\text{and}\quad
\tilde{a}<\tilde{b}\;\;\;(\text{or}\;\;\;\tilde{a}\le\tilde{b})\text{''},
\label{ineq}
\end{align}
if we decompose $a$ and $b$ into $(\alpha,\tilde a)$ and $(\beta,\tilde b)$ respectively.
This means that we can decompose $\Sigma_m(L)$ for the repetitive
quiver into the products of $\Sigma_{\tilde{m}}(L_i)$ for the
fundamental ones with different $\alpha$,
\begin{align}
\Sigma_m(L)=\sum_{s=1}^r\sum_{L_1,L_2,\cdots,L_s}F_s(r)
\prod_{i=1}^s\Sigma_{\tilde{m}}(L_i),
\end{align}
with a combinatorial factor $F_s(r)$.
Here the sum is taken over all possible partitions of $L$,
$L=L_1L_2\cdots L_s$.
The combinatorial factor $F_s(r)$ is given by counting possible
combinations of $\{\alpha_i\}_{i=1}^s$ satisfying the inequality
$1\le\alpha_1<\alpha_2<\cdots<\alpha_s\le r$,
\begin{align}
F_s(r)=\#\{(\alpha_1,\cdots,\alpha_t)|
1\le\alpha_1<\alpha_2<\cdots<\alpha_s\le r\}
=\biggl(\begin{matrix}r\\s\end{matrix}\biggr).
\end{align}
For example, the condition $1\le a\le b<c\le m$ in defining $\Sigma_m(q,p,q)$ \eqref{Sigmadef} is decomposed as
\begin{align}
(1,1)\le(\alpha,\tilde a)\le(\beta,\tilde b)
<(\gamma,\tilde c)\le(r,\tilde m),
\end{align}
where the inequalities are understood in the sense of \eqref{ineq}.
This implies that $\Sigma_m(q,p,q)$ can be decomposed into $\Sigma_{\tilde m}(q,p,q)$, $\Sigma_{\tilde m}(q,p)\Sigma_{\tilde m}(q)$, $\Sigma_{\tilde m}(q)\Sigma_{\tilde m}(p,q)$ or $\Sigma_{\tilde m}(q)^2\Sigma_{\tilde m}(p)$ respectively when $\alpha=\beta=\gamma$, $\alpha=\beta<\gamma$, $\alpha<\beta=\gamma$ or $\alpha<\beta<\gamma$.
The combinatorial factor of decomposing $\Sigma_m(q,p,q)$ into $\Sigma_{\tilde m}(q,p)\Sigma_{\tilde m}(q)$ is computed by choosing two different elements $\alpha=\beta$ and $\gamma$ out of $\{1,2,\cdots,r\}$.
In this way, we find several formulae
\begin{align}
&\Sigma_m(q)=\biggl(\begin{matrix}r\\1\end{matrix}\biggr)
\Sigma_{\tilde{m}}(q),
\nonumber\\&
\Sigma_m(q,p)=\biggl(\begin{matrix}r\\1\end{matrix}\biggr)
\Sigma_{\tilde{m}}(q,p)
+\biggl(\begin{matrix}r\\2\end{matrix}\biggr)
\Sigma_{\tilde{m}}(q)\Sigma_{\tilde{m}}(p),\\
&\Sigma_m(q,p,q)=\biggl(\begin{matrix}r\\1\end{matrix}\biggr)
\Sigma_{\tilde{m}}(q,p,q)+\biggl(\begin{matrix}r\\2\end{matrix}\biggr)
(\Sigma_{\tilde{m}}(q)\Sigma_{\tilde{m}}(p,q)
+\Sigma_{\tilde{m}}(q,p)\Sigma_{\tilde{m}}(q))
+\biggl(\begin{matrix}r\\3\end{matrix}\biggr)
\Sigma_{\tilde{m}}(q)^2\Sigma_{\tilde{m}}(p),\nonumber
\end{align}
as well as those with the role of $q$ and $p$ switched.
With these relations and \eqref{sigmasum}, one can prove that $n(0)$
in \eqref{n0} satisfies
\begin{align}
[n(0)]_r=[n(0)]_1.
\end{align}

\section{A preliminary study on ${\mathcal N}=3$ quivers}\label{N3}

Having obtained the expression of the coefficient $B$ for the
${\mathcal N}=4$ superconformal circular quiver Chern-Simons theories
in section \ref{secFS} and checked the repetition invariance in
section \ref{repetition}, in this section we shall make a digression
to comment on possible generalization of the analysis to the
${\mathcal N}=3$ cases.
It was already shown in \cite{MP} that the partition function of
${\mathcal N}=3$ Chern-Simons matrix models can also be rewritten into
that of a Fermi gas system and the sum of the perturbative terms is
given by the Airy function \eqref{Ai}.
Here the one-particle Hamiltonian of the Fermi gas system is given as
\begin{align}
e^{-{\widehat H}}=e^{-{\widehat U}_1}e^{-{\widehat U}_2}\cdots 
e^{-{\widehat U}_M},
\label{HamiltonianwhenNis3}
\end{align}
with ${\widehat U}_a$ defined by
\begin{align}
{\widehat U}_a=\log 2\cosh\frac{{\widehat P}-\nu_a{\widehat Q}}{2},
\end{align}
and $\nu_a$'s given by the Chern-Simons levels $k_a=kn_a$ as
\begin{align}
\nu_a=\sum_{b=1}^an_b,
\end{align}
which implies $\nu_M=0$.
In this section, we shall apply the analysis in section \ref{secFS} to
this theory and calculate $n(E)$ at the most leading part in $\hbar$
expansion.
Since the Hamiltonian \eqref{HamiltonianwhenNis3} is symmetric under
exchange among $\nu_a$'s at the most leading order, for the later
convenience, let us replace $\nu_a$ with $\nu_{\sigma(a)}$ so that the
new $\nu_a$ satisfies
\begin{align}
\nu_a\le \nu_{a+1},
\label{order}
\end{align}
for all $a$.
The conclusion is that the coefficients $C$ and $B$ in the Airy
function \eqref{Ai} are given by
\begin{align}
C&=\frac{2}{\pi\hbar}\sum_{a=1}^M
\frac{|\nu_{a+1}-\nu_a|}
{\sum_{b=1}^M|\nu_{a+1}-\nu_b|\sum_{c=1}^M|\nu_a-\nu_c|}
+{\cal O}(\hbar),\label{C3}\\
B&=\frac{2\pi}{3\hbar}\sum_{a=1}^M
\frac{|\nu_{a+1}-\nu_a|}
{\sum_{b=1}^M|\nu_{a+1}-\nu_b|\sum_{c=1}^M|\nu_a-\nu_c|}
-\frac{\pi}{3\hbar}\sum_{a=1}^M\frac{1}{\sum_{b=1}^M|\nu_b-\nu_a|}
+{\cal O}(\hbar),\label{B3}
\end{align}
with $\nu_{M+1}=\nu_1$

\begin{figure}
\centering
\includegraphics[width=8cm]{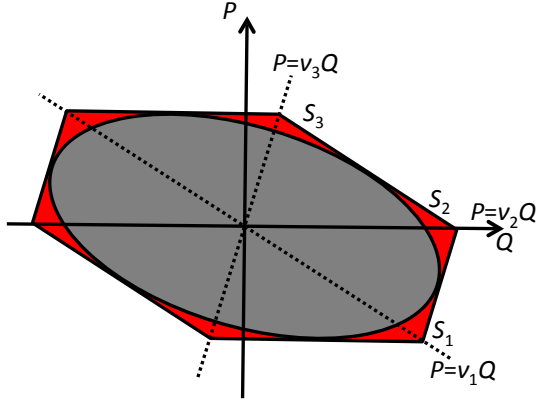}
\caption{
The Fermi surface of the ${\mathcal N}=3$ superconformal circular quiver Chern-Simons theory.
The outer polygon is the limiting convex $2M$-gon \eqref{polygon} and the inner
closed curve is the Fermi surface.}
\label{N3coloredregion}
\end{figure}

The idea of calculation is similar to the one used in section \ref{secFS} and
\cite{MP}.
At this order, the Wigner Hamiltonian is given as the classical one
\begin{align}
H_{\rm W}^{(0)}=\sum_{a=1}^MU_a,
\end{align}
with
\begin{align}
U_a=\log 2\cosh\frac{P-\nu_aQ}{2}.
\end{align}
To obtain the total volume inside the Fermi surface
\begin{align}
n(E)=\frac{1}{2\pi\hbar}
\vol\biggl\{(Q,P)\Big|
\sum_{i=a}^M\log 2\cosh\frac{P-\nu_aQ}{2}\le E\biggr\}
+{\mathcal O}(\hbar),
\label{FS3}
\end{align}
below we consider its deviation from the volume inside the convex
$2M$-gon\footnote{For simplicity, we assume the generic case
$\nu_a\ne\nu_b$ ($a\ne b$) in the following argument, though we can
justify the final results \eqref{C3} and \eqref{B3}.}
\begin{align}
\frac{1}{2\pi\hbar}
\vol\biggl\{(Q,P)\Big|\sum_{a=1}^M|P-\nu_aQ|\le 2E\biggr\},
\label{polygon}
\end{align}
where the Fermi surface \eqref{FS3} is approaching in the limit
$E\to\infty$ as in section \ref{secFS}.
(See figure \ref{N3coloredregion}.)

\begin{figure}
\centering
\includegraphics[width=10cm]{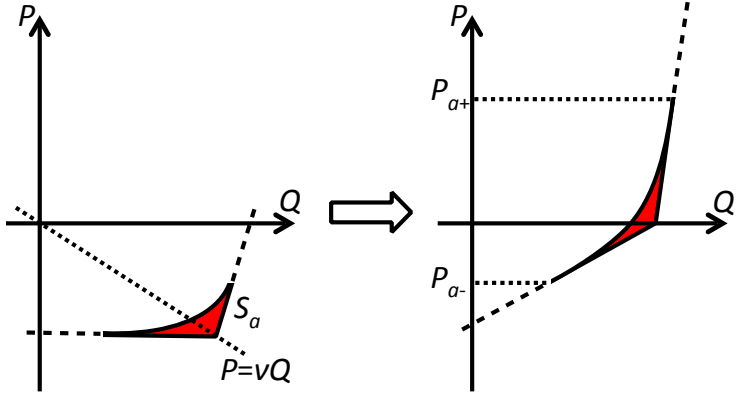}
\caption{The left figure shows the region $S_a$, the part of the
colored region, and the right figure shows its affine
transformation.}
\label{modulartrsf}
\end{figure}

Now let us calculate the volume of the deviation, the red region in
figure \ref{N3coloredregion}.
Since both $H_{\rm W}(Q,P)$ and the polygon are invariant under
$(Q,P)\rightarrow(-Q,-P)$, we can restrict ourselves to $Q>0$.
Hereafter, we shall denote as $S_a$ the region around the vertex with
$P-\nu_aQ=0$ and $Q>0$, surrounded by the curve $H_{\rm W}=E$ and the
two edges of the polygon ending on this vertex.
Since $S_a$ is distant at order $E$ from the lines $P-\nu_bQ=0$ with
$b\neq a$, on $S_a$ the Hamiltonian can be approximated up to
non-perturbative corrections in $E$ as
\begin{align}
H_{\rm W}\simeq H_{W,a}=\sum_{b(\neq a)}^M
\frac{|P-\nu_bQ|}{2}+\log 2\cosh\frac{P-\nu_aQ}{2}.
\end{align}
There are further simplification of calculation due to the invariance
of the volume under an affine transformation $(Q,P)\rightarrow
(Q,P-\nu_aQ)$ on each $S_a$.
See figure \ref{modulartrsf}.
After this affine transformation, if we denote the points on the edge
of the polygon as $(Q(P),P)$ and those on the Fermi surface $H_{W,a}=E$
as $(Q^\prime(P),P)$, we find that
\begin{align}
Q(P)-Q^\prime(P)=\frac{1}{\sum_{b=1}^M|\nu_b-\nu_a|}
\biggl(2\log 2\cosh\frac{P}{2}-|P|\biggr)+\text{non-pert}\,.
\end{align}
Therefore the volume of the region $S_a$ is
\begin{align}
\vol(S_a)=\int_{P_{a-}}^{P_{a+}}dP\frac{1}{\sum_{b=1}^M|\nu_b-\nu_a|}
\biggl(2\log 2\cosh\frac{P}{2}-|P|\biggr)+\text{non-pert}\,.
\end{align}
Here we have denoted by $P_{\pm}$ the value of the $P$-coordinate at
the midpoints of the currently considered edges of the polygon, where
the Fermi surface and the edge of polygon coalesce up to
${\cal O}(e^{-E})$. 
Since the integrand is ${\cal O}(e^{-E})$ at $|P|\sim E$, one can
extend the domain of integration to $(-\infty,\infty)$ and obtains
\begin{align}
\vol(S_a)=\frac{\pi^2}{3\sum_{b=1}^M|\nu_b-\nu_a|}+\text{non-pert}\,.
\end{align}
Subtracting them from the volume inside the polygon, one finally
obtains
\begin{align}
n(E)&=\frac{1}{2\pi\hbar}\biggl(
4E^2\sum_{a=1}^M\frac{|\nu_{a+1}-\nu_a|}
{\sum_{b=1}^M|\nu_{a+1}-\nu_b|\sum_{c=1}^M|\nu_a-\nu_c|}
-\sum_{a=1}^M\frac{2\pi^2}{3\sum_{b=1}^M|\nu_b-\nu_a|}\biggr)
\nonumber\\
&\quad+{\cal O}(\hbar)+\text{non-pert}\,.
\end{align}

If one choose the Chern-Simons levels as \eqref{pm1} so that the
supersymmetry enhances to ${\cal N}=4$, the values of $\nu_a$'s
(before rearranged as \eqref{order}) are
\begin{align}
\{\nu_a\}_{a=1}^M=\{(+1)^{q_1},(0)^{p_1},(+1)^{q_2},(0)^{p_2},
\cdots,(+1)^{q_r},(0)^{p_r}\},
\end{align}
and the classical limit of the results for ${\cal N}=4$ theories \eqref{C} and \eqref{n0} are
recovered.

Note that, although the hermiticity of the Hamiltonian is crucial in
discussing the physical Fermi surface in section \ref{secFS}, in the
${\cal N}=3$ cases the trick making Hamiltonian hermitian by unitary
transformation works only for very restricted cases.
We hope, however, to extend our results on ${\cal N}=3$ to higher
corrections in $\hbar$ by, for example, the method in section
\ref{AandJnp} in future works.

\section{WKB expansion of grand potential}
\label{AandJnp}

In this section, we shall calculate the grand potential $J(\mu)$ at
the first few leading orders in $\hbar$ including the non-perturbative
term in $\mu$.
We find that all these computations are consistent with our
perturbative result of the coefficient $B$ obtained in section
\ref{secFS}.
Besides, we have obtained several new insights on the coefficient $A$
and the non-perturbative terms, which enable us to conjecture the
expression of the coefficient $A$ \eqref{A} for the case when the
edges of $s_a=+1$ and those of $s_a=-1$ are separated, and the
expression of the first membrane instanton \eqref{c1} for the case of
$\{s_a\}_{a=1}^3=\{(+1)^2,(-1)\}$.

Again, the computation is parallel to \cite{MP}.
We write $J(\mu)$ as
\begin{align}
J(\mu)=\sum_{n=1}^\infty\frac{(-1)^{n-1}}{n}e^{\mu n}
\int\frac{dQdP}{2\pi\hbar}\bigl(e^{-n{\widehat H}}\bigr)_{\rm W},
\end{align}
and evaluate the integral for each $n$ by expanding
$(e^{-n{\widehat H}})_{\rm W}$ order by order in $\hbar$.
Then, we can substitute the results back to $J(\mu)$ and resum the series to
obtain the $\hbar$-expansion,
\begin{align}
J(\mu)=\sum_{s=0}^\infty\hbar^{s-1}J^{(s)}(\mu).
\end{align}
As stressed in \cite{MP}, there are two kind of $\hbar$ corrections to
$(e^{-n{\widehat H}})_{\rm W}$.
One is from the correction to $H_{\rm W}$ itself from
$H_{\rm W}^{(0)}$, which is partly discussed in section \ref{secFS}.
The other comes from the fact that
$H_{\rm W}\star H_{\rm W}\neq H_{\rm W}^2$.
Keeping in mind the decomposition
$e^{-n{\widehat H}}=e^{-nH_{\rm W}}e^{-n({\widehat H}-H_{\rm W})}$,
the latter contributions can be systematically treated by introducing
\begin{align}
{\mathcal G}_t=\bigl(({\widehat H}-H_{\rm W})^t\bigr)_{\rm W}.
\end{align}
The first few non-trivial examples of ${\mathcal G}_t$ are given by
\begin{align}
{\mathcal G}_2=H_{\rm W}\star H_{\rm W}-H_{\rm W}^2,\quad
{\mathcal G}_3=H_{\rm W}\star H_{\rm W}\star H_{\rm W}
-3H_{\rm W}(H_{\rm W}\star H_{\rm W})+2H_{\rm W}^3,
\quad\cdots.
\label{Grexplicit}
\end{align}
It was shown in \cite{MP} that, apart from $\hbar$ corrections to
$H_{\rm W}$ itself, the $\hbar$ expansion of ${\mathcal G}_t$ is
\begin{align}
{\mathcal G}_t=\sum_{s=2\left[\frac{t+2}{3}\right]}^\infty
\hbar^s{\mathcal G}_t^{(s)},
\label{Grexpand}
\end{align}
with ${\cal G}_t^{(s)}=0$ for any odd $s$.
With these contributions, $(e^{-n{\widehat H}})_{\rm W}$ is written as
\begin{align}
\bigl(e^{-n{\widehat H}}\bigr)_{\rm W}
=e^{-nH_{\rm W}^{(0)}}
\exp\biggl[-n\sum_{s=2}^\infty\hbar^sH_{\rm W}^{(s)}\biggr]\times
\biggl(1+\sum_{t=2}^\infty\frac{(-n)^t}{t!}{\mathcal G}_t\biggr),
\label{enhatHWexpand}
\end{align}
expanding the second and third factor, one obtains the parts which
contribute to each $J^{(s)}(\mu)$.

Below we perform these studies for $J^{(0)}(\mu)$ and $J^{(2)}(\mu)$.
Then restricting to the class of separative quivers, that is,
$\{s_a\}_{a=1}^M=\{(+1)^q,(-1)^p\}$, we calculate $J^{(4)}(\mu)$.
Note that $J^{(s)}(\mu)$ vanishes for any odd $s$ since the integrand
is always an odd function with respect to $Q$ or $P$ at this order.
In our computation the following quantity appears frequently,
\begin{align}
{\mathcal F}(a,\alpha,b,\beta,\mu)
=\sum_{n=1}^\infty\frac{(-1)^{n-1}}{n}e^{\mu n}
\int\frac{dQdP}{2\pi}
\frac{1}{\bigl(2\cosh\frac{Q}{2}\bigr)^{an+\alpha}}
\frac{1}{\bigl(2\cosh\frac{P}{2}\bigr)^{bn+\beta}}.
\end{align}
This quantity can be computed by integrating each term with the
formula
\begin{align}
\int_{-\infty}^\infty dx
\frac{1}{\bigl(2\cosh\frac{x}{2}\bigr)^{n}}
=\frac{\sqrt{4\pi}}{2^n}
\frac{\Gamma\bigl(\frac{n}{2}\bigr)}
{\Gamma\bigl(\frac{n+1}{2}\bigr)},
\label{intcosh}
\end{align}
and using the multiplication theorem of the gamma function
\begin{align}
\Gamma(mx)=\frac{m^{mx}}{\sqrt{(2\pi)^{m-1}m}}
\prod_{i=0}^{m-1}\Gamma\left(x+\frac{i}{m}\right),
\end{align}
for $m\in\mathbb{N}$ and $x\in\mathbb{R}$, so that we can use the
Pochhammer's generalized hypergeometric function
\begin{align}
\,_p\!F_q(a_1,\cdots a_p;b_1,\cdots b_q;z)
=\frac{\prod_{j=1}^q\Gamma(b_j)}{\prod_{i=1}^p\Gamma(a_i)}
\sum_{n=0}^\infty
\frac{\prod_{i=1}^p\Gamma(a_i+n)}{\prod_{j=1}^q\Gamma(b_j+n)}
\frac{z^n}{n!}.
\end{align}
Then we find that this function can be expressed as
\begin{align}
&{\mathcal F}(a,\alpha,b,\beta,\mu)
=-\frac{1}{2^{\alpha+\beta}\sqrt{ab}}\biggl[
\Bigl(\frac{-e^{\mu}}{2^{a+b}}\Bigr)^2
\frac{\Gamma(1)}{\Gamma(2)}
\prod_{i=0}^{a-1}
\frac{\Gamma(1+\frac{\alpha}{2a}+\frac{i}{a})}
{\Gamma(1+\frac{\alpha+1}{2a}+\frac{i}{a})}
\prod_{j=0}^{b-1}
\frac{\Gamma(1+\frac{\beta}{2b}+\frac{j}{b})}
{\Gamma(1+\frac{\beta+1}{2b}+\frac{j}{b})}
\nonumber\\
&\qquad\times{}_{a+b+2}F_{a+b+1}\biggl(
\Bigl\{1+\frac{\alpha}{2a}+\frac{i}{a}\Bigr\}_{i=0}^{a-1},
\Bigl\{1+\frac{\beta}{2b}+\frac{j}{b}\Bigr\}_{j=0}^{b-1},1,1;
\nonumber\\
&\qquad\qquad\qquad\qquad\quad
\Bigl\{1+\frac{\alpha+1}{2a}+\frac{i}{a}\Bigr\}_{i=0}^{a-1},
\Bigl\{1+\frac{\beta+1}{2b}+\frac{j}{b}\Bigr\}_{j=0}^{b-1},2;
\Bigl(\frac{-e^\mu}{2^{a+b}}\Bigr)^2\biggr)
\nonumber\\
&\quad-\frac{e^{\mu}}{2^{a+b}}
\frac{\Gamma(\frac{1}{2})}{\Gamma(\frac{3}{2})}
\prod_{i=0}^{a-1}
\frac{\Gamma(\frac{1}{2}+\frac{\alpha}{2a}+\frac{i}{a})}
{\Gamma(\frac{1}{2}+\frac{\alpha+1}{2a}+\frac{i}{a})}
\prod_{j=0}^{b-1}
\frac{\Gamma(\frac{1}{2}+\frac{\beta}{2b}+\frac{j}{b})}
{\Gamma(\frac{1}{2}+\frac{\beta+1}{2b}+\frac{j}{b})}
\nonumber\\
&\qquad\times{}_{a+b+2}F_{a+b+1}\biggl(
\Bigl\{\frac{1}{2}+\frac{\alpha}{2a}+\frac{i}{a}\Bigr\}_{i=0}^{a-1},
\Bigl\{\frac{1}{2}+\frac{\beta}{2b}+\frac{j}{b}\Bigr\}_{j=0}^{b-1},
\frac{1}{2},1;
\nonumber\\
&\qquad\qquad\qquad\qquad\quad
\Bigl\{\frac{1}{2}+\frac{\alpha+1}{2a}+\frac{i}{a}\Bigr\}_{i=0}^{a-1},
\Bigl\{\frac{1}{2}+\frac{\beta+1}{2b}+\frac{j}{b}\Bigr\}_{j=0}^{b-1},
\frac{3}{2};
\Bigl(\frac{-e^\mu}{2^{a+b}}\Bigr)^2\biggr)\biggr].
\label{calF}
\end{align}
In the following three subsections, we shall first compute the grand
potential order by order in $\hbar$ and express the final result using
the function ${\mathcal F}(a,\alpha,b,\beta,\mu)$.
Then, we choose several specific types of quivers $\{s_a\}_{a=1}^M$ to study the
grand potentials in the large $\mu$ expansion and guess the general behavior
of the perturbative and non-perturbative parts,
$J^{(s)}(\mu)=J^{(s)}_\text{pert}(\mu)+J^{(s)}_\text{np}(\mu)$.

\subsection{$J^{(0)}(\mu)$}
First we consider the most leading part, $J^{(0)}(\mu)$.
At this order, $(e^{-n{\widehat H}})_{\rm W}$ is simply
$e^{-nH_{\rm W}^{(0)}}$.
Since $H_{\rm W}^{(0)}$ is given as \eqref{HW0}, 
the quantity is nothing but the one computed previously\footnote{For
the $r$-ple repetition of the ABJM quiver, this result was also
obtained by Masazumi Honda by similar techniques (private note).}
\begin{align}
J^{(0)}(\mu)={\mathcal F}(\Sigma(q),0,\Sigma(p),0,\mu).
\end{align}

Studying the asymptotic behavior of $J^{(0)}(\mu)$ at
$\mu\rightarrow\infty$ for $1\le \Sigma(q)\le 4$ and $1\le\Sigma(p)\le 4$, we have
found that the perturbative part coincides with the following
expression
\begin{align}
J^{(0)}_\text{pert}(\mu)&=\frac{4\mu^3}{3\pi\Sigma(q)\Sigma(p)}
+\biggl[\frac{4\pi}{3\Sigma(q)\Sigma(p)}
-\frac{\pi}{3}
\biggl(\frac{\Sigma(p)}{\Sigma(q)}
+\frac{\Sigma(q)}{\Sigma(p)}\biggl)\biggr]\mu
\nonumber\\
&\quad+\frac{2\zeta(3)}{\pi}
\biggl(\frac{\Sigma(p)^2}{\Sigma(q)}
+\frac{\Sigma(q)^2}{\Sigma(p)}\biggr).
\label{J0pert}
\end{align}
The $\mu$ dependent part is consistent with the results obtained in
section \ref{secFS}.
We have also found that, the non-perturbative corrections consist of
terms proportional to
\begin{align}
\exp\left[-\frac{2n\mu}{\Sigma(q)}\right],\quad\text{or}\quad
\exp\left[-\frac{2n\mu}{\Sigma(p)}\right],\label{expofmbinst}
\end{align}
with $n\ge 1$ but not their bound states.
For example, for $\Sigma(q)=1,\Sigma(p)=2$ we obtain
\begin{align}
J^{(0)}_\text{np}(\mu)&=
-8e^{-\mu}+\biggl[-\frac{12\mu^2-28\mu-28}{\pi}+\pi\biggr]e^{-2\mu}
+{\mathcal O}(e^{-3\mu}),
\label{J0np21}
\end{align}
while for $\Sigma(q)=2,\Sigma(p)=3$ we find
\begin{align}
J^{(0)}_\text{np}(\mu)&=
-\frac{160\pi^2}{9\sqrt{3}\Gamma\left(-\frac{1}{3}\right)
\Gamma\left(\frac{2}{3}\right)\Gamma\left(\frac{8}{3}\right)}
e^{-\frac{2\mu}{3}}
-64e^{-\mu}
+\frac{9\cdot 2^{\frac{2}{3}}\pi^{\frac{3}{2}}}
{\Gamma\left(-\frac{5}{3}\right)\Gamma\left(\frac{7}{6}\right)}
e^{-\frac{4\mu}{3}}
+{\mathcal O}(e^{-2\mu}),
\end{align}
without e.g.\ the bound state $e^{-\frac{5}{3}\mu}$ of
$e^{-\frac{2}{3}\mu}$ and $e^{-\mu}$.

\subsection{$J^{(2)}(\mu)$}
Collecting the relevant terms in the expansion of
$(e^{-n{\widehat H}})_{\rm W}$ \eqref{enhatHWexpand}, $J^{(2)}(\mu)$
is given as
\begin{align}
J^{(2)}(\mu)=\sum_{n=1}^\infty
\frac{(-1)^{n-1}}{n}e^{\mu n}
\int\frac{dQdP}{2\pi}e^{-nH_{\rm W}^{(0)}}
\biggl[-nH_{\rm W}^{(2)}
+\frac{n^2}{2}{\mathcal G}_2^{(2)}(H^{(0)}_{\rm W})
-\frac{n^3}{6}{\mathcal G}_3^{(2)}(H^{(0)}_{\rm W})
\biggr],
\label{J2rhs}
\end{align}
where ${\mathcal G}_t^{(s)}$ is defined by \eqref{Grexplicit} and
\eqref{Grexpand}, whose several relevant terms are given explicitly by
\begin{align}
{\mathcal G}^{(2)}_2(H^{(0)}_{\rm W})
&=-\frac{1}{4}\Sigma(q)\Sigma(p)U^{\prime\prime}T^{\prime\prime},
\nonumber\\
{\mathcal G}^{(2)}_3(H^{(0)}_{\rm W})
&=-\frac{\Sigma(q)^2\Sigma(p)}{4}(U^\prime)^2T^{\prime\prime}
-\frac{\Sigma(q)\Sigma(p)^2}{4}(T^\prime)^2U^{\prime\prime}.
\end{align}
Using the integration by parts
\begin{align}
&\int dQ e^{-nH_{\rm W}^{(0)}}U^\prime g(Q,P)
=\int dQ e^{-nH_{\rm W}^{(0)}}\frac{1}{n\Sigma(q)}
\frac{\partial g}{\partial Q},\nonumber\\
&\int dP e^{-nH_{\rm W}^{(0)}}T^\prime g(Q,P)
=\int dP e^{-nH_{\rm W}^{(0)}}\frac{1}{n\Sigma(p)}
\frac{\partial g}{\partial P},
\label{byparts}
\end{align}
for an arbitrary function $g(Q,P)$, one can replace
\begin{align}
(U^\prime)^2\rightarrow\frac{1}{n\Sigma(q)}U^{\prime\prime},
\quad(T^\prime)^2\rightarrow\frac{1}{n\Sigma(p)}T^{\prime\prime},
\end{align}
in the integrand in \eqref{J2rhs}.
After these replacements, we can use our formula \eqref{calF} directly
to obtain
\begin{align}
J^{(2)}(\mu)
=\biggl[B^{(2)}-\frac{1}{24}\Sigma(q)\Sigma(p)\partial_\mu^2\biggr]
{\mathcal F}(\Sigma(q),2,\Sigma(p),2,\mu).
\end{align}

Again we calculate the asymptotic behavior of $J^{(2)}(\mu)$ for
$1\le\Sigma(q)\le 4$, $1\le\Sigma(p)\le 4$ and obtain the perturbative parts
expressed as
\begin{align}
J^{(2)}_\text{pert}(\mu)
=\frac{B^{(2)}\mu}{2\pi}
-\frac{B^{(2)}(\Sigma(q)+\Sigma(p))}{2\pi},
\label{J2pert}
\end{align}
where the term proportional to $\mu$ is consistent with the result
obtained in section \ref{secFS}.
We have also found that each term in the non-perturbative part
exhibits the same behavior \eqref{expofmbinst} as those in $J^{(0)}(\mu)$.
For $\{s_a\}_{a=1}^3=\{(+1)^2,(-1)\}$, for example, we find that
\begin{align}
J^{(2)}_\text{np}(\mu)=\frac{1}{6}e^{-\mu}
+\biggl[\frac{\mu^2-11\mu/3-1/2}{2\pi}-\frac{\pi}{24}\biggr]e^{-2\mu}
+{\mathcal O}(e^{-3\mu}).
\label{J2np21}
\end{align}
Remarkably, the exponents appearing in this expression depends only
on $(\Sigma(q),\Sigma(p))$, not on the ordering of $\{s_a\}_{a=1}^M$.
For example, for $\{s_a\}_{a=1}^4=\{(+1)^2,(-1)^2\}$, we find that
\begin{align}
J^{(2)}_\text{np}(\mu)&=
\biggl[\frac{2\mu+1}{3\pi}\biggr]e^{-\mu}
+\biggl[\frac{\mu^2-17\mu/3+7/6}{\pi}-\frac{\pi}{3}\biggr]
e^{-2\mu}+{\mathcal O}(e^{-3\mu}),
\label{++--}
\end{align}
while, for $\{s_a\}_{a=1}^4=\{(+1),(-1),(+1),(-1)\}$, we find
\begin{align}
J^{(2)}_\text{np}(\mu)&=
\biggl[-\frac{\mu^2-10\mu/3-2/3}{8\pi}+\frac{\pi}{24}\biggr]e^{-\mu}
+\biggl[\frac{5\mu^2-77\mu/3+7/6}{4\pi}
-\frac{5\pi}{12}\biggr]e^{-2\mu}+{\mathcal O}(e^{-3\mu}).
\label{+-+-}
\end{align}
Both of these last two examples share the same instanton exponents
with different polynomial coefficients.

\subsection{$J^{(4)}(\mu)$ for separative models}
The terms in \eqref{enhatHWexpand} which are relevant to
$J^{(4)}(\mu)$ are $H_{\rm W}^{(2)}$, $H_{\rm W}^{(4)}$ and
${\mathcal G}_t$ with $2\le t\le 6$.
Here we shall restrict ourselves to the case $m=1$, that is,
$\{s_a\}_{a=1}^M=\{(+1)^q,(-1)^p\}$, since $H_{\rm W}^{(4)}$ for
general circular quivers is still obscure.
In this case $H_{\rm W}^{(4)}$ is given as
\begin{align}
H_{\rm W}^{(4)}&=\frac{qp^2}{144}T^\prime T^{(3)}U^{(4)}
-\frac{q^2p}{288}U^\prime U^{(3)}T^{(4)}
-\frac{q^3p^2}{240}(U^\prime)^2U^{\prime\prime}(T^{\prime\prime})^2
+\frac{q^2p^3}{60}(T^\prime)^2T^{\prime\prime}(U^{\prime\prime})^2
\nonumber\\
&\quad-\frac{q^3p^2}{80}(U^\prime)^2U^{\prime\prime}T^\prime T^{(3)}
+\frac{q^2p^3}{120}(T^\prime)^2T^{\prime\prime}U^\prime U^{(3)}
+\frac{7q^4p}{5760}(U^\prime)^4T^{(4)}
-\frac{qp^4}{720}(T^\prime)^4U^{(4)}.
\end{align}
Though the result contains a lot of terms, it is again simplified by
using the following replacements
\begin{align}
(U^\prime)^4\rightarrow\frac{1}{(nq)^2}
\biggl(9(U^{\prime\prime})^2-\frac{3}{2}U^{\prime\prime}\biggr),\quad
(T^\prime)^4\rightarrow\frac{1}{(np)^2}
\biggl(9(T^{\prime\prime})^2-\frac{3}{2}T^{\prime\prime}\biggr),
\nonumber\\
(U^\prime)^2U^{\prime\prime}\rightarrow\frac{1}{nq}
\biggl(3(U^{\prime\prime})^2-\frac{1}{2}U^{\prime\prime}\biggr),\quad
(T^\prime)^2T^{\prime\prime}\rightarrow\frac{1}{np}
\biggl(3(T^{\prime\prime})^2-\frac{1}{2}T^{\prime\prime}\biggr),
\end{align}
which are allowed by the integrating by parts \eqref{byparts} and the
definition of $U$ and $T$ \eqref{defUT}.
One finally obtains
\begin{align}
&J^{(4)}(\mu)=\sum_{n\ge 1}\frac{(-1)^{n-1}}{n}e^{\mu n}
\int\frac{dQdP}{2\pi}e^{-nH_{\rm W}^{(0)}}\nonumber\\
&\quad\times\frac{(qp)^2(1-n^2)}{5760}\biggl[
-(9-n^2)\left((U^{\prime\prime})^2+\frac{1}{2}U^{\prime\prime}\right)
\left((T^{\prime\prime})^2+\frac{1}{2}T^{\prime\prime}\right)
+(4-n^2)U^{\prime\prime}T^{\prime\prime}\biggr].
\end{align}
After processing the integral and the sum over $n$ in the same way as
in $J^{(0)}(\mu)$ and $J^{(2)}(\mu)$, one can write $J^{(4)}(\mu)$ as
\begin{align}
J^{(4)}(\mu)&=\frac{(qp)^2}{5760}
\Bigl[-(1-\partial_\mu^2)(9-\partial_\mu^2)f_{41}
+(1-\partial_\mu^2)(4-\partial_\mu^2)f_{42}\Bigr],
\end{align}
with
\begin{align}
f_{41}&={\mathcal F}(q,4,p,4,\mu)
+\frac{1}{2}{\mathcal F}(q,2,p,4,\mu)
+\frac{1}{2}{\mathcal F}(q,4,p,2,\mu)
+\frac{1}{4}{\mathcal F}(q,2,p,2,\mu),\nonumber\\
f_{42}&={\mathcal F}(q,2,p,2,\mu).
\end{align}

We calculate its asymptotic behavior at $\mu\rightarrow\infty$ for
small $q,p$ and find that the results are consistent with the
following expression
\begin{align}
J^{(4)}_\text{pert}(\mu)=-\frac{(q+p)(qp)^2}{69120\pi}.
\label{J4pert}
\end{align}
Also, we calculate the non-perturbative effect and find
\begin{align}
J^{(4)}_\text{np}(\mu)=\frac{1}{1440}e^{-\mu}
+\biggl[-\frac{\mu^2-49\mu/15+34/15}{96\pi}+\frac{\pi}{1152}\biggr]
e^{-2\mu}
+{\mathcal O}(e^{-3\mu}),
\label{J4np21}
\end{align}
for $\{s_a\}_{a=1}^3=\{(+1)^2,(-1)\}$.

\subsection{Implication of WKB analysis}
In the above subsections we have studied the WKB expansion order by
order and guess the general form of the perturbative part of
$J^{(0)}(\mu)$, $J^{(2)}(\mu)$ and $J^{(4)}(\mu)$ for general
${\mathcal N}=4$ circular quivers.
Collecting the cubic and linear terms in $J^{(0)}(\mu)$ and
$J^{(2)}(\mu)$, it is straightforward to see that the results match
respectively with $C$ and $B$ in our Fermi surface studies in section
\ref{secFS}.
If we collect the constant terms  for the separated model from
\eqref{J0pert}, \eqref{J2pert} and \eqref{J4pert}, we find
\begin{align}
A=\frac{4\zeta(3)}{\pi}\frac{1}{2}
\biggl(\frac{p^2}{q\hbar}+\frac{q^2}{p\hbar}\biggr)
-\frac{1}{24\pi}\frac{p^2q\hbar+q^2p\hbar}{2}
-\frac{1}{34560\pi}\frac{p^2(q\hbar)^3+q^2(p\hbar)^3}{2}
+{\mathcal O}(\hbar^5).
\label{Awkb}
\end{align}
This result leads us to conjecture that the coefficient $A$ is given
in terms of that of the ABJM theory by \eqref{A}.
Also, if we collect the first instanton term for the case of
$\{s_a\}_{a=1}^3=\{(+1)^2,(-1)\}$, we find
\begin{align}
J^\text{MB}_\text{np}(\mu)&=\biggl[-\frac{4}{\pi k}+\frac{\pi k}{3}
+\frac{(\pi k)^3}{180}+{\mathcal O}(k^5)\biggr]e^{-\mu}
+\frac{1}{\pi}\biggl[\Bigl(-\frac{6}{\pi k}+\pi k-\frac{(\pi k)^3}{12}
+{\mathcal O}(k^5)\Bigr)\mu^2\nonumber\\
&\quad+\Bigl(\frac{14}{\pi k}-\frac{11\pi k}{3}+\frac{49(\pi k)^3}{180}
+{\mathcal O}(k^5)\Bigr)\mu\nonumber\\
&\quad+\Big(\frac{14}{\pi k}+\frac{\pi}{2k}
-\frac{\pi k}{2}-\frac{\pi^3k}{12}
-\frac{17(\pi k)^3}{90}+\frac{\pi^5k^3}{144}
+{\mathcal O}(k^5)\Bigr)\biggr]e^{-2\mu}
+{\mathcal O}(e^{-3\mu}).
\label{membraneinstanton}
\end{align}
This is consistent with the series expansion of \eqref{c1}.
In the next section, we shall see a strong numerical evidence for
these conjectures \eqref{A} and \eqref{c1} for the
$\{s_a\}_{a=1}^3=\{(+1)^2,(-1)\}$ case.

If we restrict ourselves to the separative case
$\{s_a\}_{a=1}^3=\{(+1)^2,(-1)\}$, we can proceed further with the
instanton expansion.
We find that the instanton takes the form
\begin{align}
J^\text{MB}_\text{np}(\mu)=\sum_{\ell=1}^\infty
\biggl[c_{2\ell-1}e^{-(2\ell-1)\mu}
+(a_{2\ell}\mu^2+b_{2\ell}\mu+c_{2\ell})e^{-2\ell\mu}\biggr].
\end{align}
As in \cite{HMO3}, we can define the functions
\begin{align}
J_a(\mu)=\sum_{\ell=1}^\infty a_{2\ell}e^{-2\ell\mu},\quad
J_b(\mu)=\sum_{\ell=1}^\infty b_{2\ell}e^{-2\ell\mu},\quad
J_c(\mu)=\sum_{\ell=1}^\infty c_\ell e^{-\ell\mu},
\end{align}
and rewrite the sum of the perturbative part and the membrane
instanton part
\begin{align}
J_\text{pert}(\mu)+\mu^2J_a(\mu)+\mu J_b(\mu)+J_c(\mu)
=J_\text{pert}(\mu_\text{eff})
+\mu_\text{eff}\widetilde J_b(\mu_\text{eff})
+\widetilde J_c(\mu_\text{eff}),
\end{align}
in terms of the effective chemical potential
\begin{align}
\mu_\text{eff}=\mu+\frac{J_a(\mu)}{C}.
\end{align}
Then we find that the two coefficients $\widetilde b_{2\ell}$ and
$\widetilde c_{2\ell}$ defined by
\begin{align}
\widetilde J_b(\mu_\text{eff})
=\sum_{\ell=1}^\infty\widetilde b_{2\ell}e^{-2\ell\mu_\text{eff}},
\quad
\widetilde J_c(\mu_\text{eff})
=\sum_{\ell=1}^\infty\widetilde c_{\ell}e^{-\ell\mu_\text{eff}},
\end{align}
satisfy the derivative relation
\begin{align}
\widetilde c_{2\ell}=-k^2\frac{\partial}{\partial k}
\frac{\widetilde b_{2\ell}}{2\ell k}.
\end{align}
We have checked it for $1\le\ell\le 4$.
This structure \cite{HMO3} was important in the ABJM case for the result to be
expressed in terms of the refined topological string \cite{HMMO}.
This makes us to expect the theory to be solved as in the ABJM case.

\section{Cancellation mechanism beyond ABJM}\label{CancelMech}

In the previous sections, we have studied mainly the perturbative part
of the general ${\mathcal N}=4$ superconformal circular quiver
Chern-Simons theories.
Here we shall look more carefully into the non-perturbative effects by
restricting ourselves to a certain model.
Aside from the ABJM matrix model, which has a dual description of the
topological string theory on local ${\mathbb P}^1\times{\mathbb P}^1$,
the next-to-simplest case would probably be the separated one with
$\{s_a\}_{a=1}^3=\{(+1)^2,(-1)\}$.
We shall see explicitly the first sign that this theory has a similar
interesting structure in the instanton expansion.
Namely, both the coefficients of the worldsheet instanton and the
membrane instanton contain poles at certain coupling constants, though
the poles are cancelled in the sum.
First, let us note that the membrane instanton effect of this model has been fixed to
be \eqref{c1} in \eqref{membraneinstanton} and is divergent when $k$
is an even number $k=k_\text{even}$,
\begin{align}
J^\text{MB}_\text{np}(\mu)\sim-\frac{4}{\pi(k-k_\text{even})}e^{-\mu}.
\label{MBdiv}
\end{align}
Hereafter, we shall see that the divergence at $k=2$ is cancelled by the first worldsheet instanton. 

We also determine the total non-perturbative effects by following the
strategy of \cite{HMO2}.
We first compute the exact values of the partition function $Z(N)$ up
to a certain number $N_\text{max}$ \cite{HMO1,PY,HMO2}.
We have computed them for
$(k,N_\text{max})=(1,20),(2,13),(3,7),(4,9),(5,3),(6,7)$.
Several examples are listed in table \ref{exact}.
\begin{table}
\begin{align*}
&Z_2(1)=\frac{1}{8\pi},\quad
Z_2(2)=\frac{-8+\pi^2}{1024\pi^2},\quad
Z_2(3)=\frac{-600+61\pi^2}{368640\pi^3},\quad
Z_2(4)=\frac{960-9424\pi^2+945\pi^4}{94371840\pi^4},\nonumber\\
&\quad 
Z_2(5)=\frac{2479680-1928080\pi^2+169899\pi^4}{237817036800\pi^5},
\nonumber\\
&\quad
Z_2(6)=\frac{14999040+110004160\pi^2-118324488\pi^4+10843875\pi^6}
{91321742131200\pi^6},\nonumber\\
&Z_3(1)=\frac{1}{12\pi},\quad
Z_3(2)=\frac{-864+89\pi^2}{31104\pi^2},\quad
Z_3(3)=\frac{-21384+13311\pi^2-2048\sqrt{3}\pi^3}{10077696\pi^3},
\nonumber\\
&\quad 
Z_3(4)=\frac{614304-1821312\pi^2-32768\sqrt{3}\pi^3+196297\pi^4}
{1934917632\pi^4},\nonumber\\
&\quad
Z_3(5)=\frac{339072480-997174800\pi^2+44236800\sqrt{3}\pi^3
+936266499\pi^4-158617600\sqrt{3}\pi^5}{15672832819200\pi^5},
\nonumber\\
&\quad
Z_3(6)=(-5845063680+55396185120\pi^2+530841600\sqrt{3}\pi^3
-110714929056\pi^4\nonumber\\
&\qquad\qquad\qquad-2124840960\sqrt{3}\pi^5+11796983935\pi^6)
/(2708265511157760\pi^6),\quad
\nonumber\\
&Z_4(1)=\frac{1}{16\pi},\quad
Z_4(2)=\frac{-48+5\pi^2}{8192\pi^2},\quad
Z_4(3)=\frac{-2640+833\pi^2-180\pi^3}{5898240\pi^3},\nonumber\\
&\quad
Z_4(4)=\frac{6400-15776\pi^2-4864\pi^3+3081\pi^4}{402653184\pi^4},
\nonumber\\
&\quad Z_4(5)=\frac{48625920-83759200\pi^2+11894400\pi^3
+38045661\pi^4-10773000\pi^5}
{30440580710400\pi^5},\nonumber\\
&\quad
Z_4(6)=(-1157345280+10549584640\pi^2+5902848000\pi^3
-17773668432\pi^4
\nonumber\\
&\qquad\qquad\qquad-9397728000\pi^5+4494764925\pi^6)
/(46756731971174400\pi^6),\nonumber\\
&Z_5(1)=\frac{1}{20\pi},\quad 
Z_5(2)=\frac{-7000+(3145-1088\sqrt{5})\pi^2}{400000\pi^2},
\nonumber\\
&\quad Z_5(3)=\frac{-300000+(367025-14400\sqrt{5})\pi^2
-18432\sqrt{50-10\sqrt{5}}\pi^3}{360000000\pi^3},\nonumber\\
&Z_6(1)=\frac{1}{24\pi},\quad
Z_6(2)=\frac{-3240+331\pi^2}{746496\pi^2},\quad
Z_6(3)=\frac{-495720+287037\pi^2-43520\sqrt{3}\pi^3}
{2418647040\pi^3},\nonumber\\
&\quad
Z_6(4)=\frac{459794880-1161396144\pi^2
-320716800\sqrt{3}\pi^3+289774225\pi^4}
{50153065021440\pi^4},\nonumber\\
&\quad
Z_6(5)=(572595791040-1548287349840\pi^2+122276044800\sqrt{3}\pi^3
+1331543069217\pi^4\nonumber\\
&\qquad\qquad\qquad-229345715200\sqrt{3}\pi^5)
/(1137471514686259200\pi^5),
\nonumber\\
&\quad
Z_6(6)=(-9765317657088+73750628879424\pi^2
+30831120875520\sqrt{3}\pi^3\nonumber\\
&\qquad\qquad\qquad
-143992509769800\pi^4-81529317310464\sqrt{3}\pi^5
+57069728465365\pi^6)
\nonumber\\
&\qquad\qquad\qquad
/(786220310951142359040\pi^6).
\end{align*}
\caption{Exact values of the partition function $Z_k(N)$ of the model $\{s_a\}_{a=1}^3=\{(+1)^2,(-1)\}$.}
\label{exact}
\end{table}

Then, we assume the polynomial expression for the instanton
coefficient in the grand potential to be the same form as that in the
ABJM case and fit the data of the exact values in table \ref{exact}
with the corresponding expression of the partition function to find out
the unknown coefficients.
We can then determine the coefficients from those with larger
contribution in $\mu\to\infty$ one by one.
For example, if the grand potential is given by
\begin{align}
J^{k=4}(\mu)=\frac{C}{3}\mu^3+B\mu+A
+\gamma_1e^{-\frac{1}{2}\mu}
+(\alpha_2\mu^2+\beta_2\mu+\gamma_2)e^{-\mu}
+\gamma_3e^{-\frac{3}{2}\mu}
+{\mathcal O}(e^{-2\mu}),
\end{align}
We fit the exact values of $Z(N)$ against the function
\begin{align}
&Z(N)=e^AC^{-1/3}\biggl(\Ai\Bigl[C^{-1/3}(N-B)\Bigr]
+\gamma_1\Ai\Bigl[C^{-1/3}\Bigl(N+\frac{1}{2}-B\Bigr)\Bigr]
\nonumber\\
&\quad+\Bigl(\alpha_2\partial_N^2-\beta_2\partial_N+\gamma_2
+\frac{1}{2}\gamma_1^2\Bigr)
\Ai\Bigl[C^{-1/3}\Bigl(N+1-B\Bigr)\Bigr]\nonumber\\
&\quad+\Bigl(\gamma_3+\gamma_1
\bigl(\alpha_2\partial_N^2-\beta_2\partial_N+\gamma_2\bigr)
+\frac{1}{6}\gamma_1^3\Bigr)
\Ai\Bigl[C^{-1/3}\Bigl(N+\frac{3}{2}-B\Bigr)\Bigr]
\biggr),
\end{align}
with the six unknown coefficients $A$, $\gamma_1$, $\alpha_2$,
$\beta_2$, $\gamma_2$, $\gamma_3$.
We can first confirm the coincidence between the numerical value of
$A$ and our expected value of $A$ \eqref{A}.
After that, we plug in the expected exact value \eqref{A} and repeat
the same fitting to determine $\gamma_1$.
Note that, unlike the ABJM matrix model, since the exponential decay
is rather slow, we find a better accuracy if we include coefficients
of the higher instanton effects into fitting.

Finally we find that, from the numerical studies of the partition
function of the separative model with
$\{s_a\}_{a=1}^3=\{(+1)^2,(-1)\}$, the grand potential is given by
\begin{align}
&J^{k=2}_\text{np}(\mu)=\biggl[\frac{2\mu+2}{\pi}\biggr]e^{-\mu}
+{\mathcal O}(e^{-2\mu}),\quad
J^{k=3}_\text{np}(\mu)=\frac{8}{3}e^{-\frac{2}{3}\mu}
+{\mathcal O}(e^{-\frac{4}{3}\mu}),\nonumber\\
&J^{k=4}_\text{np}(\mu)=2\sqrt{2}e^{-\frac{1}{2}\mu}
+{\mathcal O}(e^{-\mu}),\quad
J^{k=5}_\text{np}(\mu)=\frac{8}{\sqrt{5}}e^{-\frac{2}{5}\mu}
+{\mathcal O}(e^{-\frac{4}{5}\mu}),\nonumber\\
&J^{k=6}_\text{np}(\mu)=\frac{8}{\sqrt{3}}e^{-\frac{1}{3}\mu}
+{\mathcal O}(e^{-\frac{2}{3}\mu}).
\label{Jk}
\end{align}
The comparison of these exact values with the numerical values
obtained from fitting can be found in table \ref{fit}.
Note that, although we only display the first several exact values of
the partition function in table \ref{exact}, we have used our full set
of exact values in obtaining the numerical values.
Aside from the case of $k=5$ where we have only a few data, as a whole
we find a very good match.

\begin{table}[t]
\begin{center}
\begin{tabular}{c|c|r|r}
&&numerical values& expected exact values\\
\hline\hline
$k=2$&$A$
&$-0.3103048519$&$-0.3103048520$\\
&$\alpha_1$
&$0.0000001660$&$0\phantom{.0000000000}$\\
&$\beta_1$
&$0.6366192705$&$2/\pi\simeq 0.6366197724$\\
&$\gamma_1$
&$0.6366205469$&$2/\pi\simeq 0.6366197724$\\
\hline\hline
$k=3$&$A$
&$-0.8115986816$&$-0.8115986811$\\
&$\gamma_1$
&$2.666666666$&$8/3\simeq 2.666666667$\\
\hline\hline
$k=4$&$A$
&$-1.368159992$&$-1.368159992$\\
&$\gamma_1$
&$2.828426464$&$2\sqrt{2}\simeq 2.828427125$\\
\hline\hline
$k=5$&$A$
&$-2.014160398$&$-2.014179117$\\
&$\gamma_1$
&$3.577692778$&$8/\sqrt{5}\simeq 3.577708764$\\
\hline\hline
$k=6$&$A$
&$-2.762757648$&$-2.762757648$\\
&$\gamma_1$
&$4.618802104$&$8/\sqrt{3}\simeq 4.618802154$
\end{tabular}
\end{center}
\caption{Comparison of numerical values obtained from fitting and
expected exact values for the perturbative coefficient $A$ and the
non-perturbative ones.
The expected exact values for $A$ is given in \eqref{A} written in
terms of the ABJM value \eqref{aABJM} while the expect values for the
first instanton effects are given in \eqref{Jk}.}
\label{fit}
\end{table}

Since there are no other contributions than the worldsheet instanton
in the first instanton effect in $J^{k=3,4,5,6}(\mu)$, we expect that
these coefficients should be explained by the first worldsheet
instanton.
We find a good interpolating function for it as in \eqref{d1}.
Note that this factor is divergent at integers $k=1,2$.
At $k=2$, we find that 
\begin{align}
J^\text{WS}_\text{np}(\mu)\sim
\biggl[\frac{4}{\pi(k-2)}+\frac{2(\mu+1)}{\pi}\biggr]e^{-\mu},
\end{align}
where the divergence cancels completely with \eqref{MBdiv} which is
coming from the membrane instanton \eqref{c1} and the finite part
reproduces the numerical study \eqref{Jk}.
This is a non-trivial consistency check of our conjecture of
\eqref{c1} and \eqref{d1}.

\section{Discussion}\label{Discussion}

In this paper we have studied the partition functions of
superconformal Chern-Simons theories of the circular quiver type using
the Fermi gas formalism.
Aside from the preliminary study in section \ref{N3}, our main target
is the cases where the supersymmetry is enhanced to ${\mathcal N}=4$.
Following the argument that the perturbative part should sums up to
the Airy function \eqref{Ai} in this case as well, we have explicitly
determined the perturbative coefficient $B$ \eqref{B} for the general
${\mathcal N}=4$ cases.
We also find a conjectural form \eqref{A} of the coefficient $A$
for the special case where the two colors of edges in the circular
quiver diagram are separated, i.e.\ $\{s_a\}_{a=1}^M=\{(+1)^q,(-1)^p\}$.
We further restrict ourselves to the case of
$\{s_a\}_{a=1}^3=\{(+1)^2,(-1)\}$, which is the simplest case
next to the ABJM case, and study the non-perturbative effects.
We find that the non-perturbative effects enjoy the similar
cancellation mechanism as in the ABJM case.
Both the coefficients of the worldsheet instanton and those of the membrane
instanton are divergent at certain levels, though the divergences are
cancelled completely.

We would like to stress that our study is one of the first signals that
it is possible to generalize the success in the ABJM theory to more
general theories whose relation with the topological string theory is
not so clear.
Namely, after finding out that for the ABJM theory the cancellation of
divergences in coefficients \cite{HMO2} helps to determine the grand
potential in terms of the refined topological string theory
on local $\mathbb{P}^1\times\mathbb{P}^1$ \cite{HMMO}, the ABJ theory
\cite{HLLLP2,ABJ} was studied carefully in \cite{AHS,H,MM,HO,HNS,K}
using its relation to the topological string theory
\cite{MPtop,DMP1}.
Here for the general ${\mathcal N}=4$ superconformal theories of the
circular quiver type, though the direct relation to the topological
strings is still unclear, our study suggests that most of the methods
used in the ABJ(M) theory are also applicable.
The final result may correspond to some deformations of the
topological string theories and \cite{HW} may be helpful along this
line.

We hope to extend the results on the ABJM theories to the class of
models with $\{s_a\}_{a=1}^M=\{(+1)^q,(-1)^p\}$,
which we believe that it is appropriate to call the ``$(q,p)$-minimal
model'' in ${\cal N}=3$ quiver Chern-Simons theories.
Even more, maybe we can finally solve all of the ${\mathcal N}=4$ or
${\mathcal N}=3$ Chern-Simons theories and understand the whole moduli
space by studying the cancellation mechanism among various instanton
effects. 

Before it, there are many basic points to be fixed firstly.
For example, in this paper we have restricted ourselves to the
theories with hermitian Hamiltonians in the Fermi gas formalism.
We believe, however, that our result \eqref{B} works for the non-hermitian cases to some extent by the following two observations.
First, the result \eqref{n0} from the Fermi surface analysis is consistent with that from the WKB analysis \eqref{J2pert} where we do not refer to the hermiticity.
Second, the formal expression associated to the non-hermitian higher commutators reduces finally to vanishing non-perturbative terms \eqref{higher}.
It is desirable to give a more concrete argument for the non-hermitian cases.
Also, though we have given a few non-trivial evidences for our
conjecture of the coefficient $A$ for the separative models, it is
desirable to prove it rigorously and write down a formula for the
general case.

In this paper we have displayed the coefficients of the membrane
instanton \eqref{c1} and the worldsheet instanton \eqref{d1} for the
next-to-simplest (2,1) separative model, $\{s_a\}_{a=1}^3=\{(+1)^2,(-1)\}$.
Actually we can continue to the coefficients of higher instantons.
We can find an exact function expression for the second membrane
instanton coefficient which is consistent with the WKB expansion in
section \ref{AandJnp}.
Also, we can repeat the numerical fitting in section \ref{CancelMech}
to higher instantons as in the ABJM case \cite{HMO2,HMO3} to find an
exact function expression for the second and third worldsheet
instanton coefficients.
It seems that the cancellation mechanism works as well.
However, we decide not to display them because the evidences are not
enough yet.

It is also interesting to observe that the $k=1$ and $k=2$ grand potentials in the (2,1) model resemble respectively to the $k=2$ grand potential in the ABJM theory and that in \cite{HaOk} with $N_f=4$. 
This implies that in general the $k=1$ grand potential in the $(2q,1)$ model is related to the $k=2$ grand potential in the $(q,1)$ model with the signs of the odd instanton terms reversed.
Using the results in \cite{HaOk}, we have checked this relation also for $q=3,4,6$.

Obviously it is interesting to reproduce many of our prediction from
the gravity side.
Let us list several discussions.
First we have seen the shift of the coefficient $B$ \eqref{B}, which implies the
shift of the 't Hooft coupling constant
\begin{align}
\lambda_{\rm eff}=\lambda-B^{(2)}-\frac{B^{(0)}}{k^2}.
\label{tHooftshift}
\end{align}
We would like to see its origin in the gravity dual along the line of
\cite{AHHS}.
Next the result of the WKB expansion \eqref{expofmbinst} implies that the membrane instanton can wrap
on the Lagrangian submanifolds which have the volume divided by the
factors of $\Sigma(q)$ and $\Sigma(p)$.
It would be interesting to reproduce these effects from the gravity
dual.
It was known \cite{IY} that the ordering of \eqref{sqp} corresponds to
the extra discrete torsion in the orbifold background.
In this sense, we find it natural that this effect appears only in the
shift of the 't Hooft coupling and in the coefficient polynomials as in
\eqref{++--} and \eqref{+-+-}.
We would like to understand this effect better.
Along the line of the interpretation in the gravity dual, it is very
interesting to note that the coefficient of the one-loop log term was
studied from the gravity side \cite{BGMS} and the match with the
expansion of the Airy function \eqref{Ai} was found.
Also, very recently, the Airy function was reproduced from the
localization computation in the gauged supergravity \cite{DDG}.

Finally, though we have used the matrix model \eqref{CSMM} obtained
after localization for the partition functions of superconformal
Chern-Simons theories, it would be interesting to study the
non-perturbative instanton effects directly from the field-theoretical
viewpoints.

\section*{Acknowledgements}
We are grateful to Yasuyuki Hatsuda, Shinji Hirano, Masazumi Honda,
Yosuke Imamura, Marcos Mari\~no, Fuminori Nakata, Kazumi Okuyama,
Chaiho Rim, Masaki Shigemori, Asuka Takatsu for valuable discussions.
The work of S.M.\ is supported by JSPS Grant-in-Aid for Scientific
Research (C) \# 26400245, while the work of T.N.\ is partly supported
by the JSPS Research Fellowships for Young Scientists.
The final stage of this work was done when S.M.\ was attending the
conference ``Finite-size Technology in Low Dimensional Quantum Systems
(VII)'', which was supported in part by JSPS Japan-Hungary Research
Cooperative Program.

\end{document}